\documentclass[11pt]{article}

\usepackage[margin=1.25in]{geometry}

\usepackage{microtype}
\usepackage{graphicx}
\usepackage{subfigure}
\usepackage{booktabs} %
\usepackage{placeins}

\usepackage{hyperref}

\usepackage{amsmath}
\usepackage{amssymb}
\usepackage{mathtools}
\usepackage{amsthm}

\usepackage[capitalize,noabbrev]{cleveref}

\usepackage{amsfonts}
\usepackage{tikz}
\usepackage{tikz-cd}
\usetikzlibrary{decorations.pathreplacing} %
\usepackage{natbib}
\usepackage{svg}
\usepackage{enumerate}
\usepackage{paralist}
\usepackage{multirow}
\usepackage{authblk}
\usepackage{algorithm}
\usepackage{algorithmic}
\usepackage{natbib}

\newcommand{\IP}{\mathcal{I}}
\newcommand{\MIP}{\mathcal{MI}}
\DeclareMathOperator{\ICP}{ICP}
\DeclareMathOperator{\AS}{IAS}
\DeclareMathOperator{\PA}{PA}
\DeclareMathOperator{\CH}{CH}
\DeclareMathOperator{\DE}{DE}
\DeclareMathOperator{\AN}{AN}

\DeclareMathOperator{\MB}{MB}
\newcommand{\indep}{\mbox{${}\perp\mkern-11mu\perp{}$}}

\newcommand{\G}{\mathcal{G}}
\DeclareMathOperator{\gene}{\texttt{gene}}
\newcommand{\IASest}{IAS$_\text{est.\ graph}$}

\usepackage[normalem]{ulem}

\theoremstyle{plain}
\newtheorem{theorem}{Theorem}[section]
\newtheorem{proposition}[theorem]{Proposition}

\theoremstyle{definition}
\newtheorem{definition}[theorem]{Definition}
\newtheorem{assumption}[theorem]{Assumption}
\theoremstyle{remark}
\newtheorem{remark}[theorem]{Remark}

\begin{document}
\title{Invariant Ancestry Search}
\date{}
\author[1]{Phillip B. Mogensen\thanks{pbm@math.ku.dk}}
\author[1]{Nikolaj Thams}
\author[1]{Jonas Peters}
\affil[1]{Department of Mathematical Sciences, University of Copenhagen, Denmark}
\maketitle

\begin{abstract}
Recently, methods have been proposed that exploit the invariance of prediction models with respect to changing environments to infer subsets of the causal parents of a response variable. If the environments influence only few of the underlying mechanisms, the subset identified by invariant causal prediction (ICP), for example, may be small, or even empty. We introduce the concept of minimal invariance and propose invariant ancestry search (IAS). In its population version, IAS outputs a set which contains only ancestors of the response and is a superset of the output of ICP. When applied to data, corresponding guarantees hold asymptotically if the underlying test for invariance has asymptotic level and power. We develop scalable algorithms and perform experiments on simulated and real data. 
\end{abstract}

\section{Introduction}
Causal reasoning addresses the challenge of understanding why systems behave the way they do and what happens if we actively intervene. 
Such mechanistic understanding is inherent to human cognition, and developing statistical methodology that learns and utilizes causal relations is a key step in improving both narrow and broad AI \citep{jordan2019artificial,pearl2018theoretical}.
Several approaches exist for learning causal structures from observational data. Approaches such as the PC-algorithm \citep{spirtes2000causation} or greedy equivalence search \citep{chickering2002optimal} learn (Markov equivalent) graphical representations of the causal structure \cite{lauritzen1996graphical}. 
Other approaches learn the graphical structure under additional assumptions, such as non-Gaussianity \cite{shimizu2006linear} or non-linearity \cite{hoyer2008nonlinear, Peters2014jmlr}. \citet{zheng2018dags} convert the problem into a continuous optimization problem, at the expense of identifiability guarantees. 

Invariant causal prediction (ICP) \citep{peters2016causal,heinze2018invariant,pfister2019invariant,gamella2020active,martinet2021variance} assumes that data are sampled from heterogeneous environments (which can be discrete, categorical or continuous), and identifies direct causes of a target $Y$, also known as causal parents of $Y$.
Learning ancestors (or parents) of a response $Y$ yields understanding of anticipated changes when intervening in the system. It is a less ambitious task than learning the complete graph but may allow for methods that come with weaker assumptions and stronger guarantees.
More concretely, for predictors $X_1, \ldots, X_d$, ICP searches for subsets $S\subseteq \{1, \ldots, d\}$ that are invariant; a set $X_S$ of predictors is called invariant if 
it renders $Y$ independent of the environment,
conditional on $X_S$.
ICP then outputs the intersection of all invariant predictor sets $S_{\ICP}\coloneqq \cap_{S\,\text{invariant}} S$. \citet{peters2016causal} show that if invariance is tested empirically from data at level $\alpha$, the resulting intersection $\hat{S}_{\ICP}$ is a subset of direct causes of $Y$ with probability at least $1-\alpha$.%
\footnote{\citet{rojas2018invariant, Magliacane2018, arjovsky2019invariant, christiansen2021causal} propose techniques that consider similar invariance statements with a focus on distribution generalization instead of causal discovery.}

In many cases, however, the set learned by ICP forms a strict subset of all direct causes or may even be empty. This is because disjoint sets of predictors can be invariant, yielding an empty intersection, which may happen both for finite samples 
as well as in the population setting.
In this work, we introduce and characterize minimally invariant sets of predictors, that is, invariant sets $S$ for which no proper subset is invariant. We propose to consider the union $S_{\AS}$ of all minimally invariant sets, where IAS stands for invariant ancestry search. We prove that $S_{\AS}$ is a subset of causal ancestors of $Y$, invariant, non-empty and contains $S_{\ICP}$. 
Learning causal ancestors of a response may be desirable for several reasons:
e.g., they are the variables that may have an influence on the response variable when intervened on.
In addition, because IAS yields an invariant set, it can be used to construct predictions that are stable across environments \citep[e.g.,][]{rojas2018invariant,christiansen2021causal}.

In practice, we  estimate minimally invariant sets using a test for invariance. If such a test has asymptotic power against some of the non-invariant sets (specified in \cref{sec:AS}), we show that, asymptotically, the probability of $\hat{S}_{\AS}$ being a subset of the ancestors is at least $1 - \alpha$. This puts stronger assumptions on the invariance test than ICP (which does not require any power) in return for discovering a larger set of causal ancestors.
We prove that our approach retains the ancestral guarantee if we test minimal invariance only among subsets up to a certain size. 
This yields a computational speed-up compared to testing minimal invariance in all subsets, but comes at the cost of potentially finding fewer causal ancestors.

The remainder of this work is organized as follows. 
In \cref{sec:preliminaries} we review relevant background material, and we introduce the concept of minimal invariance in \cref{sec:minimal-invariance}. \Cref{sec:oracle-algorithm} contains an oracle algorithm for finding minimally invariant sets 
(and a closed-form expression of $S_{\ICP}$)
and \cref{sec:5} presents theoretical guarantees when testing minimal invariance from data. 
In \cref{sec:experiments} we evaluate our method in several simulation studies as well as a real-world data set on gene perturbations. Code is provided at \url{https://github.com/PhillipMogensen/InvariantAncestrySearch}.

\section{Preliminaries}\label{sec:preliminaries}
\subsection{Structural Causal Models and Graphs}
We consider a setting where data are sampled from a structural causal model (SCM) \cite{pearl2009causality,bongers2021foundations}
\begin{equation*}
    Z_j \coloneqq f_j(\PA_j, \epsilon_j),
\end{equation*}
for some functions $f_j$, parent sets $\PA_j$ and noise distributions $\epsilon_j$. 
Following \cite{peters2016causal,heinze2018invariant}, we consider an SCM over variables $Z \coloneqq (E, X, Y)$ where $E$ is an exogenous environment variable (i.e., $\PA_{E} = \emptyset$), $Y$ is a response variable and $X = (X_1, \ldots, X_d)$ is a collection of predictors of $Y$. We denote by $\mathcal{P}$ the family of all possible distributions induced by an SCM over $(E, X, Y)$ of the above form.

For a collection of nodes $j \in [d]\coloneqq \{1, \ldots, d\}$ and their parent sets $\PA_j$, we define a directed graph $\G$ with nodes $[d]$
 and  edges $j' \to j$ for all $j'\in \PA_j$. 
 We denote by $\CH_j$, $\AN_j$ and $\DE_j$ the children, ancestors and descendants of a variable $j$, respectively, neither containing~$j$. A graph $\G$ is called a directed acyclic graph (DAG) if it does not contain any directed cycles.
See \citet{pearl2009causality} for more details and the definition of $d$-separation.

Throughout the remainder of this work, we make the following assumptions about causal sufficiency and exogeneity of $E$ (\cref{sec:extensions} describes how these assumptions can be relaxed). 
\begin{assumption}\label{assumption:1}
    Data are sampled from 
    an SCM over nodes $(E, X, Y)$, such that the corresponding graph is a DAG, the distribution is faithful with respect to this DAG, and the environments are exogenous,
    i.e., $\PA_E = \emptyset$.
\end{assumption}

\subsection{Invariant Causal Prediction}\label{sec:ICP}
Invariant causal prediction (ICP), introduced by 
\citet{peters2016causal}, exploits the existence of heterogeneity in the data, here encoded by an environment variable $E$, to learn a subset of causal parents of a response variable $Y$. 
A subset of predictors $S \subseteq [d]$ is \emph{invariant} if $Y \indep E \mid S$, and we define $\IP \coloneqq \{S \subseteq [d] \mid S \text{ invariant}\}$ to be the set of all invariant sets.
We denote the corresponding hypothesis that $S$ is invariant by
\begin{equation*}
    H_{0,S}^{\IP}: \quad S \in \IP.
\end{equation*}
Formally, $H_{0, S}^{\IP}$ corresponds to a subset of distributions in $\mathcal{P}$, and we denote by $H_{A, S}^{\IP} := \mathcal{P} \setminus H_{0, S}^{\IP}$ the alternative hypothesis to $H_{0, S}^{\IP}$.
\citet{peters2016causal} define the oracle output 
\begin{equation}\label{eq:S_ICP}
S_{\ICP} \coloneqq \bigcap_{S: H_{0, S}^{\IP} \, \text{true}} S
\end{equation}
(with $S_{\ICP} = \emptyset$ if no sets are invariant)
and prove $S_{\ICP} \subseteq \PA_Y$.
If provided with a test for 
the hypotheses 
$H_{0, S}^{\IP}$,
we can test all sets $S \subseteq [d]$ for invariance and take the intersection over all accepted sets: $\hat S_{\ICP} \coloneqq \bigcap_{S: H_{0,S}^{\IP}\,\text{not rejected}} S$; 
If the invariance test has level $\alpha$, 
$\hat S_{\ICP} \subseteq \PA_Y$ 
with probability at least $1-\alpha$. 

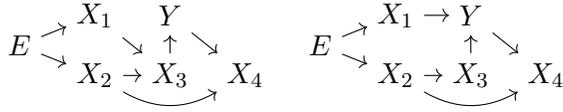
\begin{figure}
    \centering
    \begin{tikzpicture}
        \node (E) at (0, 0) {$E$};
        \node (X1) at (1, 0.4) {$X_1$};
        \node (X2) at (1, -0.4) {$X_2$};
        \node (X3) at (2, -0.4) {$X_3$};
        \node (Y) at (2, 0.4) {$Y$};
        \node (X4) at (3, -0.4) {$X_4$};
        \draw[->] (E) to (X1);
        \draw[->] (E) to (X2);
        \draw[->] (X1) to (X3);
        \draw[->] (X2) to (X3);
        \draw[->] (X2) to[bend right=30] (X4);
        \draw[->] (X3) to (Y);
        \draw[->] (Y) to (X4);
        \node (E) at (4, 0) {$E$};
        \node (X1) at (5, 0.4) {$X_1$};
        \node (X2) at (5, -0.4) {$X_2$};
        \node (X3) at (6, -0.4) {$X_3$};
        \node (Y) at (6, 0.4) {$Y$};
        \node (X4) at (7, -0.4) {$X_4$};
        \draw[->] (E) to (X1);
        \draw[->] (E) to (X2);
        \draw[->] (X1) to (Y);
        \draw[->] (X2) to (X3);
        \draw[->] (X2) to[bend right=30] (X4);
        \draw[->] (X3) to (Y);
        \draw[->] (Y) to (X4);
    \end{tikzpicture}
    \caption{Two structures where $S_{\ICP}\subsetneq \PA_Y$. (\textit{left}) $S_{\ICP} = \emptyset$. (\textit{right}) $S_{\ICP} = \{1\}$. In both, our method outputs $S_{\AS} = \{1, 2, 3\}$.}
    \label{fig:two-bad-structures}
\end{figure}

However, even for the oracle output in \cref{eq:S_ICP}, there are many graphs for which $S_{\ICP}$ is a strict subset of $\PA_Y$. For example, in 
\cref{fig:two-bad-structures} (left), since both $\{1, 2\}$ and $\{3\}$ are invariant, $S_{\ICP} \subseteq \{1, 2\} \cap \{3\} = \emptyset$. This does not violate $S_{\ICP}\subseteq \PA_Y$, but is non-informative. Similarly, in \cref{fig:two-bad-structures} (right), $S_{\ICP}=\{1\}$, as all invariant sets contain $\{1\}$.
Here, $S_{\ICP}$ contains some information, but is not able to recover the full parental set.
In neither of these two cases, $S_{\ICP}$ is an invariant set.
If the environments are such that each parent of $Y$ is either affected by the environment directly or is a parent of an affected node, then $S_{\ICP} = \PA_Y$ \citep[proof of Theorem~3]{peters2016causal}. The shortcomings of ICP thus relate to settings where the environments act on too few variables or on uninformative ones.

For large $d$, %
it has been suggested to apply ICP to the variables in the \emph{Markov boundary} \cite{PearlJudea2014Prii}, $\MB_Y = \PA_Y \cup \CH_Y \cup \PA(\CH_Y)$ %
(we denote the oracle output by $S_{\ICP}^{\MB}$).
As $\PA_Y\subseteq \MB_Y$, it still holds that $S_{\ICP}^{\MB}$ is a subset of the causal parents of the response.\footnote{In fact, $S_{\ICP}^{\MB}$ is always at least as informative as ICP. E.g., there exist graphs in which $S_{\ICP} = \emptyset$ and $S_{\ICP}^{\MB} \neq \emptyset$, see \cref{fig:two-bad-structures} (left). There are no possible structures for which $S_{\ICP}^{\MB} \subsetneq S_{\ICP}$, as both search for invariant sets over all sets of parents of $Y$.} However, the procedure must still be applied to $2^{\vert\MB_Y\vert}$ sets, which is only feasible if the Markov boundary is sufficiently small.
In practice, the Markov boundary can, for example, be estimated using Lasso regression or gradient boosting techniques \citep{Tibshirani96,meinshausen2006high,friedman2001greedy}. 

\section{Minimal Invariance and Ancestry}\label{sec:minimal-invariance}
We now introduce the concept of minimally invariant sets, which are invariant sets that do not have any invariant subsets. 
We propose to consider $S_{\AS}$, the oracle outcome of invariant ancestry search, defined as the union of all minimally invariant sets. We will see that $S_{\AS}$ is an invariant set, it consists only of ancestors of $Y$, and it contains $S_{\ICP}$ as a subset.

\begin{definition}\label{definition:MIP}
Let $S \subseteq [d]$. We say that $S$ is \emph{minimally invariant} if and only if 
    \begin{equation*}
    S \in \IP \,\, \text{and} \,\, 
    \forall S' \subsetneq S: \,\, S' \not\in \IP;
    \end{equation*}
that is, 
$S$ is invariant and no subset of $S$ is invariant.
We define $\MIP \coloneqq \{S \mid S \text{ minimally invariant}\}$.
\end{definition}
The concept of minimal invariance is closely related to the concept of minimal $d$-separators 
\citep{tian1998finding}. This connection allows us to state several properties of minimal invariance. For example, an invariant set is minimally invariant if and only if it is non-invariant as soon as one of its elements is removed. 
\begin{proposition}
\label{theorem:simpleDef}
Let $S \subseteq [d]$. 
Then $S \in \MIP$ if and only if 
 $S \in \IP$ and for all $j \in S$, it holds that $S \setminus \{j\} \not\in \IP$.
\end{proposition}
The proof follows directly from \citep[][Corollary 2]{tian1998finding}.
We can therefore decide whether 
a given invariant set $S$
is minimally invariant using $\mathcal{O}(\vert S\vert)$ checks for invariance, rather than $\mathcal{O}(2^{\vert S\vert})$ (as suggested by Definition~\ref{definition:MIP}). We use this insight in \cref{sec:MIP_tests}, when we construct a 
statistical test
for whether or not a set is minimally invariant.

To formally define the oracle outcome of IAS, 
we denote the hypothesis that a set $S$ is minimally invariant by
    \begin{equation*}
        H_{0, S}^{\MIP}: \quad S \in \MIP
    \end{equation*}
(and the alternative hypothesis, $S \notin \MIP$, by $H_{A, S}^{\MIP}$)
and define the quantity of interest
    \begin{equation}\label{eq:S_AS}
        S_{\AS} \coloneqq \bigcup_{S: H_{0, S}^{\MIP} \, \text{true}} S
    \end{equation}
with the convention that a union over the empty set is the empty set. 

The following proposition states that $S_{\AS}$ is a subset of the ancestors of the response $Y$. Similarly to $\PA_Y$, variables in $\AN_Y$ are causes of $Y$ in that for each ancestor there is a directed causal path to $Y$. Thus, generically, when intervened, these variables have a causal effect on the response.
\begin{proposition}\label{prop:tian1998}
    It holds that $S_{\AS} \subseteq \AN_Y$.
\end{proposition}
The proof follows directly from \citep[][Theorem~2]{tian1998finding}; see also \citep[][Proposition~2]{acid2013algorithm}. The setup in these papers is more general than what we consider here; we therefore provide direct proofs 
for Propositions~\ref{theorem:simpleDef} and~\ref{prop:tian1998} 
in \cref{appendix:proofs}, which may provide further intuition for the results.

Finally, we show that the oracle output of IAS contains that of ICP and, contrary to ICP, 
it is always an invariant set.
\begin{proposition}\label{theorem:MIP_properties}
Assume that $E \not\in \PA_Y$. It holds that
\begin{compactitem}
    \item[(i)] $S_{\AS} \in \IP$ and 
    \item[(ii)] $S_{\ICP}\subseteq S_{\AS}$, with equality if and only if $S_{\ICP}\in \IP$.
\end{compactitem}
\end{proposition}

\section{Oracle Algorithms}\label{sec:oracle-algorithm}
When provided with an oracle that tells us whether a set is invariant or not, how can we efficiently compute $S_{\ICP}$ and $S_{\AS}$? 
Here, we assume that the oracle is given by a DAG, see \cref{assumption:1}.
A direct application of Equations~\eqref{eq:S_ICP} and~\eqref{eq:S_AS} would require checking a number of sets that grows exponentially in the number of nodes. 
For $S_{\ICP}$, we have the following characterization.\footnote{To the best of our knowledge, this characterization is novel.}
\begin{proposition}\label{lemma:S_ICP}
If $E \not\in \PA_Y$, then
$S_{\ICP} = \PA_Y \cap \left(\CH_E \cup \PA(\AN_Y\cap\CH_E)\right)$. 
\end{proposition}
This allows us to efficiently read off $S_{\ICP}$ from the DAG, (e.g., it can naively be done in  $\mathcal{O}((d+2)^{2.373}\log(d+2))$ time, where the exponent $2.373$ comes from matrix multiplication).
For $S_{\AS}$, to the best of our knowledge, there is no closed form expression that has a similarly simple structure. 

Instead, for IAS, we exploit the recent development of efficient algorithms for computing all minimal $d$-separators (for two given sets of nodes) in a given DAG \citep[see, e.g.,][]{tian1998finding,van2019separators}.
A set $S$ is called a \emph{minimal $d$-separator} of $E$ and $Y$ if it
$d$-separates $E$ and $Y$ given $S$ and no strict subset of $S$ satisfies this property.
These algorithms are often motivated by determining minimal adjustment sets \citep[e.g.,][]{pearl2009causality} that can be used to compute the total causal effect between two nodes, for example. If the underlying distribution is Markov and faithful with respect to the DAG, then a set $S$ is minimally invariant if and only if it is a minimal $d$-separator for $E$ and $Y$. We can therefore use
the same algorithms to find minimally invariant sets;
\citet{van2019separators}
provide an algorithm (based on work by \citet{Takata2010}) for finding minimal $d$-separators 
with polynomial delay time.
Applied to our case, this means that while there may be exponentially many minimally invariant sets,\footnote{This is the case if there are $d/2$ (disjoint) directed paths 
between $E$ and $Y$, with each path containing two $X$-nodes, for example \citep[e.g.,][]{van2019separators}.} when listing all such sets it takes at most polynomial time until the next set or the message that there are nor further sets is output. 
In practice, on random graphs, we found this to work well 
(see Section~\ref{experiment:1}).
But since $S_{\AS}$ is the union of all minimally invariant sets, even faster algorithms may be available;
to the best of our knowledge, it is an open question whether finding $S_{\AS}$ is an NP-hard problem (see Appendix~\ref{appendix:minimal_review} for details).

We provide a function for listing all minimally invariant sets in our python code; it uses an implementation of the above mentioned algorithm, provided in the R \citep{Rsoftware} package \texttt{dagitty} \citep{dagitty}.
In \cref{experiment:1}, we study the properties of the oracle set $S_{\AS}$. When applied to $500$ randomly sampled, dense graphs with $d = 15$ predictor nodes and five interventions, the \texttt{dagitty} implementation had a median speedup of 
a factor of roughly $17$, compared to a brute-force search (over the ancestors of $Y$). The highest speedup achieved was by a factor of more than $1{,}900$.

The above mentioned literature can be used only for oracle algorithms, where the graph is given. In the following sections, we discuss how to test the hypothesis of minimal invariance from data. 

\section{Invariant Ancestry Search} \label{sec:5}
\subsection{Testing a Single Set for Minimal Invariance}\label{sec:MIP_tests}
Usually, we neither observe a full SCM nor its graphical structure.
Instead, we observe data from an SCM, which we want to use to decide whether a set is in $\MIP$, such that we make the correct decision with high probability.
We now show that a set $S$ can be tested for minimal invariance with asymptotic level and power if given a test for invariance that has asymptotic level and power.

Assume that $\mathcal{D}_n = (X_i, E_i, Y_i)_{i = 1}^n$ are 
observations (which may or may not be independent) of $(X, E, Y)$ and let $\phi_n^{\MIP}:
\operatorname{powerset}([d]) \times \mathcal{D}_n \times (0, 1) \to \{0, 1\}$ be a decision rule 
that transforms $(S, \mathcal{D}_n, \alpha)$
 into a decision $\phi_n^{\MIP}(S, \mathcal{D}_n, \alpha)$ about whether the hypothesis $H_{0, S}^{\MIP}$ should be rejected ($\phi_n^{\MIP} = 1$) at significance threshold $\alpha$, or not ($\phi_n^{\MIP} = 0$). 
To ease notation, we suppress the dependence on $\mathcal{D}_n$ and $\alpha$ 
when the statements are unambiguous.

A test $\psi_n$ for the hypothesis $H_0$
has pointwise asymptotic level if 
    \begin{equation}\label{eq:level}
        \forall \alpha \in (0, 1): \quad
         \sup\limits_{\mathbb{P} \in H_0}\lim\limits_{n \rightarrow \infty} \mathbb{P} (\psi_n  = 1) \leq \alpha
    \end{equation}
    and pointwise asymptotic power if
    \begin{equation}\label{eq:power}
        \forall \alpha \in (0, 1): \quad
         \inf\limits_{\mathbb{P} \in H_{A}}   \lim\limits_{n \to \infty} \mathbb{P} (\psi_n  = 1) = 1.
    \end{equation}
    If the limit and  the supremum (resp. infimum) in \cref{eq:level} (resp. \cref{eq:power}) can be interchanged, we say that $\psi_n$ has uniform asymptotic level (resp. power).

Tests for invariance have been examined in the literature. \citet{peters2016causal} propose two simple methods for testing for invariance in linear Gaussian SCMs when the environments are discrete, although the methods proposed extend directly to other regression scenarios. \citet{pfister2019invariant} propose resampling-based tests for sequential data from linear Gaussian SCMs.
Furthermore, any valid test for conditional independence between $Y$ and $E$ given a set of predictors $S$ can be used to test for invariance. Although for continuous $X$, there exists no general conditional independence test that has both level and non-trivial power \citep{shah2020hardness}, it is possible to impose restrictions on the data-generating process that ensure the existence of non-trivial tests \citep[e.g.,][]{Fukumizu2008, Zhang2011uai, berrett2020conditional, shah2020hardness,thams2021statistical}. 
\citet{heinze2018invariant} provide an overview and a comparison of several  conditional independence tests in the context of invariance.

To test whether a set $S \subseteq [d]$ is minimally invariant, we define the decision rule
        \begin{equation}\label{eq:phiMIP}
            \phi_n^{\MIP} (S) \coloneqq 
                \begin{cases}
                    1 & \text{if } \phi_n(S) = 1 \,\, \text{or} \,\, \min\limits_{j \in S} \phi_n(S \setminus \{j\}) = 0,\\
                0 & \text{otherwise,}
                \end{cases}
        \end{equation}
where $\phi_n^{\MIP}(\emptyset) \coloneqq \phi_n(\emptyset)$. 
Here, $\phi_n$ is a test for the hypothesis $H_{0,S}^{\mathcal{I}}$, e.g., one of the tests mentioned above.
This decision rule rejects $H_{0, S}^{\MIP}$ either if $H_{0,S}^{\IP}$ is rejected by $\phi_n$ or if there exists $j\in S$ such that $H_{0, S\setminus \{j\}}^{\IP}$ is not rejected. 
If $\phi_n$ has pointwise (resp. uniform) asymptotic level and power, then $\phi_n^{\MIP}$ has pointwise (resp. uniform) asymptotic level and pointwise (resp. uniform) asymptotic power of at least $1 - \alpha$.

\begin{theorem}\label{theorem:MIP_test}
    Let $\phi_n^{\MIP}$ be defined as in \cref{eq:phiMIP} and let $S \subseteq [d]$. Assume that the decision rule $\phi_n$ has pointwise asymptotic level and power %
    for $S$ and for all $S \setminus \{j\},  j \in S$.
    Then, $\phi_n^{\MIP}$ has pointwise asymptotic level
    and pointwise asymptotic power of at least $1 - \alpha$, i.e.,
        \begin{equation*}
        \inf\limits_{\mathbb{P} \in H_{A, S}^{\MIP}}
            \lim_{n \rightarrow \infty} \mathbb{P}
            (\phi_n^{\MIP}(S) = 1) \geq 1 - \alpha.
        \end{equation*}
    
    If $\phi_n$ has uniform asymptotic level and power, %
    then $\phi_n^{\MIP}$ has uniform asymptotic level
    and uniform asymptotic power of at least $1 - \alpha$.
\end{theorem}

Due to \cref{prop:tian1998}, a test for $H_{0, S}^{\MIP}$ is implicitly a test for $S \subseteq \AN_Y$, and can thus be used to infer whether intervening on $S$ will have a potential causal effect on $Y$. However, rejecting $H_{0, S}^{\MIP}$ is not evidence for $S \not\subseteq \AN$; it is evidence for $S \not\in \MIP$.  

\subsection{Learning $S_{\AS}$ from Data}
\label{sec:AS}
We now consider the task of estimating the set $S_{\AS}$ from data. %
If we are given a test for invariance that has asymptotic level and power and if we correct for multiple testing appropriately, we can estimate $S_{\AS}$ by $\hat{S}_{\AS}$, which, asymptotically, is a subset of $\AN_Y$ with large probability.
\begin{theorem}\label{theorem:exhaustive_search}
Assume that the decision rule $\phi_n$ 
has pointwise asymptotic level for all minimally invariant sets and pointwise asymptotic power for all $S \subseteq [d]$ such that $S$ is not a superset of a minimally invariant set.
Define $C:=2^d$ and let
    $
        \widehat{\IP} \coloneqq \left\{
            S \subseteq [d] \mid
                \phi_n(S, \alpha C^{-1}) = 0)
        \right\}
    $
    be the set of all sets for which the hypothesis of invariance is not rejected and define
        $
        \widehat{\MIP} \coloneqq \left\{
            S \in \widehat{\IP} \mid
                \forall S' \subsetneq S: S' \not\in \widehat{\IP}
        \right\}
        $
      and
        $
        \hat{S}_{\AS} \coloneqq \bigcup_{S \in \widehat{\MIP}} S.
        $
It then holds that
    \begin{align*}
        \lim\limits_{n \to \infty} \mathbb{P}(\hat{S}_{\AS} \subseteq \AN_Y) 
        & \geq \lim\limits_{n \to \infty} \mathbb{P}(\hat{S}_{\AS} = S_{\AS}) \\
        & \geq 1 - \alpha.
    \end{align*}
\end{theorem}
A generic algorithm for implementing $\hat{S}_{\AS}$ is given in \cref{appendix:S_AS_algorithm}.

\begin{remark}\label{remark:alpha0}
    Consider a decision rule $\phi_n$ that just (correctly) rejects the empty set (e.g., because  the $p$-value is just below the threshold $\alpha$),
    indicating that the effect of the environments is weak.
    It  is likely that there are other sets $S' \not\in\IP$, which the test may not have sufficient power against and are (falsely) accepted as invariant. 
If one of such sets contains non-ancestors of $Y$, this yields a violation of $\hat{S}_{\AS} \subseteq \AN_Y$.
    To guard against this, testing $S = \emptyset$ can be done at a lower significance level, $\alpha_0 < \alpha$. This modified IAS approach is conservative and may return $\hat{S}_{\AS} = \emptyset$ if the environments do not have a strong impact on $Y$, but it retains the guarantee $\lim_{n\rightarrow\infty} \mathbb{P}(\hat{S}_{\AS} \subseteq \AN_Y) \geq 1 - \alpha$ of \cref{theorem:exhaustive_search}. 
\end{remark}

The multiple testing correction performed in \cref{theorem:exhaustive_search}
is strictly conservative because we only need to correct for the number of minimally invariant sets, and there does not exist $2^d$ minimally invariant sets.
Indeed,
the statement of 
\cref{theorem:exhaustive_search} remains valid for $C=C'$ if the underlying DAG has at most $C'$ minimally invariant sets. 
We hypothesize that 
a DAG can contain at most $3^{\lceil d/3 \rceil}$ minimally invariant sets and therefore propose using $C = 3^{\lceil d/3 \rceil}$ in practice.
If this hypothesis is true, 
\cref{theorem:exhaustive_search} remains valid (for any DAG), using $C = 3^{\lceil d/3 \rceil}$
(see \cref{appendix:simulate_max} for a more detailed discussion).

Alternatively, as shown in the following section, we can restrict the search for minimally invariant sets to a predetermined size.
This requires milder correction factors and
comes with computational benefits.

\subsection{Invariant Ancestry Search in 
Large Systems}
\label{sec:higher-dimensions}
We now develop a variation of \cref{theorem:exhaustive_search}, which allows us to search for ancestors of $Y$ in large graphs, at the cost of only identifying minimally invariant sets up to some a priori determined size.

Similarly to ICP (see Section~\ref{sec:ICP}), %
one could restrict IAS to the variables in $\MB_Y$ but the output may be smaller than $S_{\AS}$; in particular, there are only non-parental ancestors in $\MB_Y$ if these are parents to both a parent a child of $Y$
(For instance, in the graph $E \rightarrow X_1 \rightarrow \ldots \rightarrow X_d \rightarrow Y$, $S_{\AS} = \{1, \ldots, d\}$ but restricting IAS to $\MB_Y$ would yield the set $\{d\}$.)
Thus, we
do not expect such an approach to be particularly fruitful in learning ancestors.

Here, we propose an alternative approach and define
    \begin{equation}\label{eq:reducedS_IAS}
        S_{\AS}^m \coloneqq \bigcup\limits_{S: S \in \MIP \,\, \text{and} \,\, \vert S \vert \leq m } S
    \end{equation}
as the union of minimally invariant sets that are no larger than $m \leq d$. For computing $S_{\AS}^m$, one only needs to check invariance of the $\sum_{i = 0}^{m} \binom{d}{i}$ sets that are no larger than $m$. %
$S_{\AS}^m$ itself, however, can be larger than $m$: in the graph above \cref{eq:reducedS_IAS}, $S_{\AS}^1 = \{1, \dots, d\}$.
The following proposition characterizes properties of $S_{\AS}^m$.
\begin{proposition}\label{prop:smallMIP_properties}
Let $m < d$ and let $m_{\min}$ and $m_{\max}$ be the size of a smallest and a largest minimally invariant set, respectively. The following statements are true:
    \begin{compactitem}
        \item[(i)] $S_{\AS}^m \subseteq \AN_Y$.
        \item[(ii)] If $m \geq m_{\max}$, then $S_{\AS}^m = S_{\AS}$.
        \item[(iii)] If $m \geq m_{\min}$ and $E \not\in \PA_Y$, then $S_{\AS}^m \in \IP$.
        \item[(iv)] If $m \geq m_{\min}$ and $E \not\in \PA_Y$, then $S_{\ICP} \subseteq S_{\AS}^m$ with equality if and only if $S_{\ICP} \in \IP$. 
    \end{compactitem}
\end{proposition}
If $m < m_{\min}$ and $S_{\ICP} \neq \emptyset$, then $S_{\ICP} \subseteq S_{\AS}^m$ does not hold.
However, we show  in \cref{experiment:1} using simulations that $S_{\AS}^m$ is larger than $S_{\ICP}$ in many sparse graphs, even for $m = 1$, when few nodes are intervened on.

In addition to the computational speedup offered by considering $S_{\AS}^m$ instead of $S_{\AS}$, the set $S_{\AS}$ can be estimated from data using a smaller correction factor than the one employed in \cref{theorem:exhaustive_search}. This has the benefit that in practice, smaller sample sizes may be needed to  detect non-invariance.

\begin{theorem}\label{theorem:restricted_search}
Let $m \leq d$ and define $C(m) \coloneqq \sum_{i = 0}^{m} \binom{d}{i}$.
Assume that the decision rule $\phi_n$ has pointwise asymptotic level for all minimally invariant sets of size at most $m$ and pointwise power for all sets of size at most $m$ that are not supersets 
of a minimally invariant set. Let 
$\widehat{\IP}^m \coloneqq \left\{S \subseteq [d] \mid \phi_n(S, \alpha C(m)^{-1}) = 0  \,\, \text{and} \,\, \vert S \vert \leq m \right\}$, 
be the set of all sets of size at most $m$ for which the hypothesis of invariance is not rejected and define 
$\widehat{\MIP}^{m} \coloneqq \left\{S \in \widehat{\IP}^m \mid \forall S' \subsetneq S: S' \not\in \widehat{\IP}^m\right\}$
and
$\hat{S}_{\AS}^{m} \coloneqq \bigcup_{S \in \widehat{\MIP}^{m}} S$.
It then holds that
    \begin{align*}
        \lim\limits_{n \to \infty} \mathbb{P}(\hat{S}_{\AS}^{m} \subseteq \AN_Y) 
        & \geq \lim\limits_{n \to \infty} \mathbb{P}(\hat{S}_{\AS}^{m} = S_{\AS}^m) \\
        & \geq 1 - \alpha.
    \end{align*}
\end{theorem}
The method proposed in \cref{theorem:restricted_search} outputs a non-empty set if there exists a non-empty set of size at most $m$, for which the hypothesis of invariance cannot be rejected.
In a sparse graph, it is likely that many small sets are minimally invariant, whereas if the graph is dense, it may be that all invariant sets are larger than $m$, such that $S_{\AS}^m = \emptyset$.
In dense graphs however, many other approaches may fail too; for example, it is also likely that the size of the Markov boundary is so large that applying $\ICP$ on $\MB_Y$ is not feasible.

\section{Experiments}\label{sec:experiments}
We apply the methods developed in this paper in a population-case experiment using oracle knowledge (Section~\ref{experiment:1}), a synthetic experiment using finite sample tests (Section~\ref{sec:experiment2}), and a real-world data set from a gene perturbation experiment (Section~\ref{sec:experiment_gene}). In \cref{experiment:1,sec:experiment2} we consider a setting with two environments: an observational environment ($E = 0$) and an intervention environment  ($E = 1$), and examine how the strength and number of interventions affect the performance of IAS.

\subsection{Oracle IAS in Random Graphs}
\label{experiment:1}
For the oracle setting, we know that $S_{\AS} \subseteq \AN_Y$ (Proposition~\ref{prop:tian1998}) and
$S_{\ICP} \subseteq S_{\AS}$ 
(Proposition~\ref{theorem:MIP_properties}).
We first verify that the inclusion $S_{\ICP} \subseteq S_{\AS}$ 
is often strict in low-dimensional settings when there are few interventions. Second, we show that the set $S_{\AS}^m$ is often strictly larger than the set $S_{\ICP}^{\MB}$ in large, sparse graphs with few interventions.

In principle, for a given number of covariates, one can enumerate all DAGs and, for each DAG, compare $S_{\ICP}$ and $S_{\AS}$. However, because the space of DAGs grows super-exponentially in the number of nodes \citep{chickering2002optimal}, this is infeasible.
Instead, we sample graphs from the space of all DAGs that satisfy \cref{assumption:1} and $Y \in \DE_E$ (see \cref{appendix:graph_sampling} for details).

In the low-dimensional setting ($d \leq 20$), we compute $S_{\ICP}$ and $S_{\AS}$, whereas in the larger graphs ($d \geq 100$), we compute $S_{\ICP}^{\MB}$ and the reduced set $S_{\AS}^{m}$ for $m \in \{1, 2\}$ when $d = 100$ and for $m = 1$ when $d = 1{,}000$. Because there is no guarantee that 
IAS outputs a superset of ICP when searching only up to sets of some size lower than $d$, 
we compare the size of the sets output by either method. For the low-dimensional setting, we consider both sparse and dense graphs, but for larger dimensions, we only consider sparse graphs.
In the sparse setting, the DAGs are constructed such that there is an expected number of $d+1$ edges between the $d+1$ nodes $X$ and $Y$; in the dense setting, the expected number of edges equals $0.75 \cdot d (d + 1) / 2$.

\begin{figure}[tb]
    \centerline{%
    \includegraphics[width=0.7\linewidth]{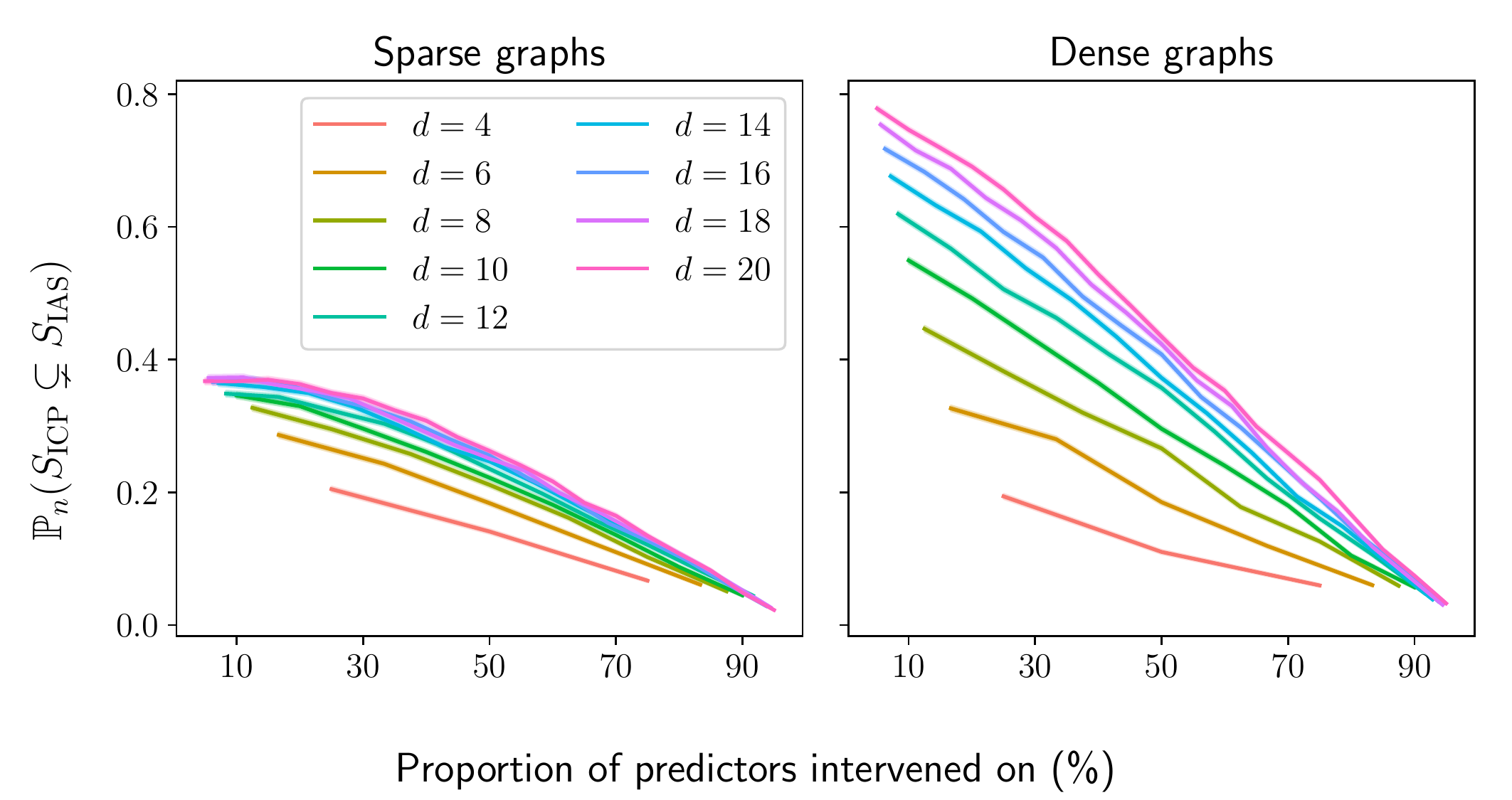}
    }
    \caption{%
    Low-dimensional oracle experiment, see \cref{experiment:1}. In all cases, 
    as predicted by theory, $S_{\ICP}$ is contained in $S_{\AS}$. 
    For many graphs, $S_{\AS}$ is strictly larger than $S_{\ICP}$. On average, this effect is more expressed when there are fewer intervened nodes. 
    $\mathbb{P}_n$ refers to the distribution used to sample graphs and every point in the figure is based on $50{,}000$ independently sampled graphs;
    $d$ denotes the number of covariates $X$. Empirical confidence bands are plotted around each line, but are very narrow.}\label{fig:experiment1_fig1}
\end{figure}

\begin{figure}[tb]
    \centerline{%
    \includegraphics[width=0.7\linewidth]{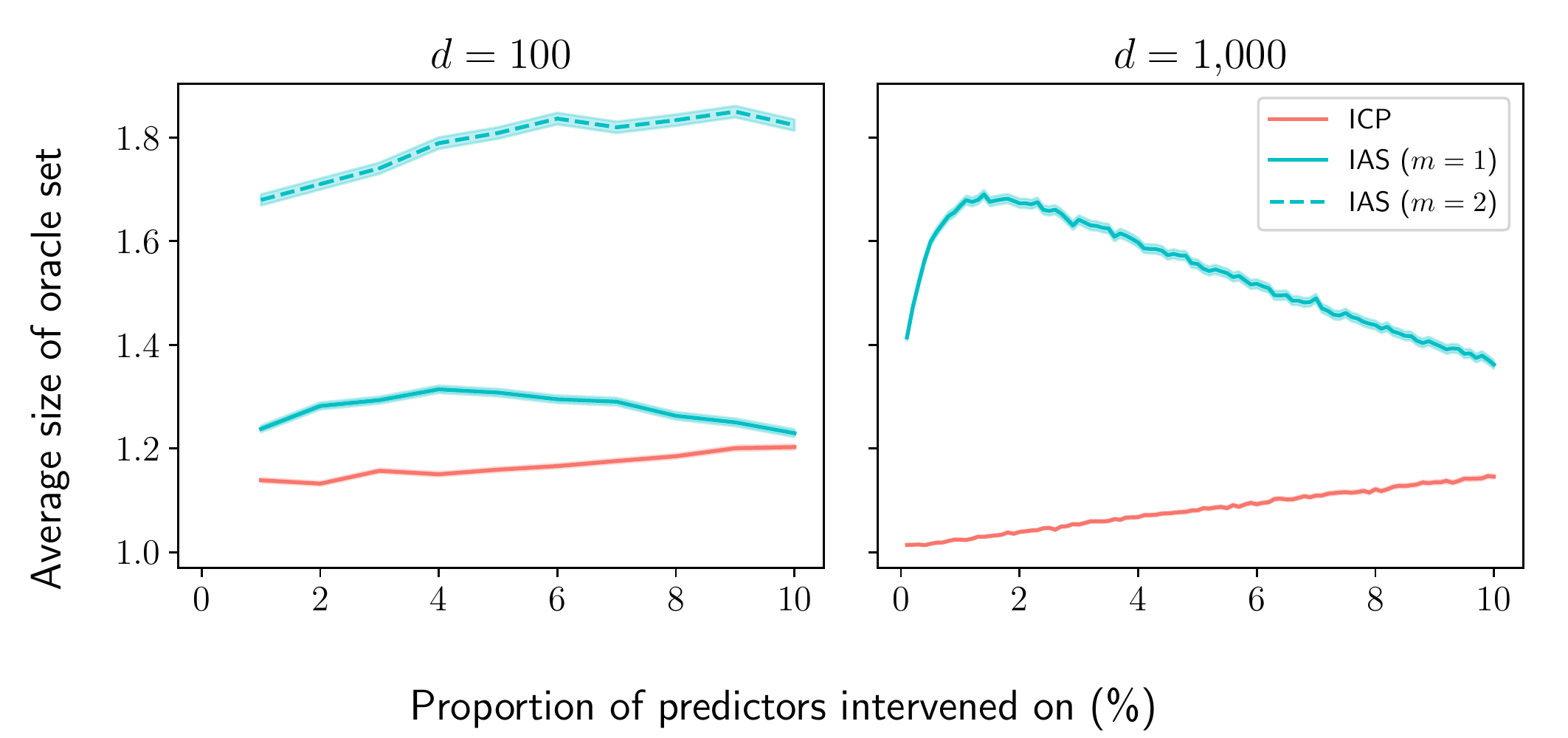}
    }  
    \caption{%
    High-dimensional oracle experiment with sparse graphs, see \cref{experiment:1}.
    The average size of the set $S_{\AS}^m$ is larger than the average size of the set $S_{\ICP}^{\MB}$,
    both when using $\AS$ to search for sets up to sizes $m = 1$ and $m = 2$. Except for the choice of $d$, the setup is the same as in \cref{fig:experiment1_fig1}.}\label{fig:experiment1_fig2}
\end{figure}

The results of the simulations are displayed in \cref{fig:experiment1_fig1,fig:experiment1_fig2}. 
In the low-dimensional setting, 
$S_{\AS}$ is a strict superset of $S_{\ICP}$
for many graphs. This effect is the more pronounced, the larger the $d$ and the fewer nodes are intervened on, see Figure~\ref{fig:experiment1_fig1}.
In fact, when there are interventions on all predictors, we know that $S_{\AS} = S_{\ICP} = \PA_Y$ \citep[][Theorem~2]{peters2016causal}, and thus the probability that $S_{\ICP} \subsetneq S_{\AS}$ is exactly zero. 
For the larger graphs, we find that the set $S_{\AS}^m$ is, on average, larger than $S_{\ICP}^{\MB}$, in particular when $d = 1{,}000$ or when $m = 2$, see Figure~\ref{fig:experiment1_fig2}. 
In the setting with $d = 100$ and $m = 1$, the two sets are roughly the same size, when $10\%$ of the predictors are intervened on. The set $S_{\ICP}^{\MB}$ becomes larger than $S_{\AS}^1$ after roughly $15\%$ of the predictors nodes are intervened on (not shown).
For both $d = 100$ and $d = 1{,}000$, the average size of the Markov boundary of $Y$ was found to be approximately $3.5$. %


\subsection{Simulated Linear Gaussian SCMs}\label{sec:experiment2}

In this experiment, we show through simulation that IAS finds more ancestors than ICP in a finite sample setting when applied to linear Gaussian SCMs.
To compare the outputs of IAS and ICP, we use the \emph{Jaccard similarity} between $\hat{S}_{\AS}$ ($\hat{S}_{\AS}^1$ when $d$ is large) and $\AN_Y$, and between $\hat{S}_{\ICP}$ ($\hat{S}_{\ICP}^{\hat{\MB}}$ when $d$ is large\footnote{$\hat{\MB}$ is a Lasso regression estimate of $\MB_Y$ containing at most 10 variables}) and $\AN_Y$.\footnote{The Jaccard similarity between two sets $A$ and $B$ is defined as $J(A, B) \coloneqq \vert A \cap B \vert / \vert A \cup B \vert$, with $J(\emptyset, \emptyset) = 0$. The Jaccard similarity equals one if the two sets are equal, zero if they are disjoint and takes a value in $(0, 1)$ otherwise.}

We sample data from sparse linear Gaussian models with i.i.d.\ noise terms in two scenarios, $d=6$ and $d=100$. In both cases, coefficients for the linear assignments are drawn randomly. We consider two environments; one observational and one interventional; in the interventional environment, we apply do-interventions of strength one to children of $E$, i.e., we fix the value of a child of $E$ to be one.
 We standardize the data along the causal order, to prevent variance accumulation along the causal order \citep{reisach2021beware}. Throughout the section, we consider a significance level of $\alpha = 5\%$. For a detailed description of the simulations, see \cref{app:simulation_details}.

To test for invariance, we employ the test used in \citet{peters2016causal}:
We calculate a $p$-value for the hypothesis of invariance of $S$ by first
linearly regressing $Y$ onto $X_S$ (ignoring $E$), and second testing whether the mean and variance of the prediction residuals is equal across environments. For details, see \citet[][Section 3.2.1]{peters2016causal}. 
\citet{schultheiss2021} also consider the task of estimating ancestors 
but since their method is uninformative for Gaussian data and does not consider environments, it is not directly applicable here.

In \cref{theorem:exhaustive_search}, we assume asymptotic power of our invariance test. 
When $d = 6$, we test hypotheses with a correction factor $C = 3^{\lceil 6 /3 \rceil} = 9$, as suggested in \cref{appendix:simulate_max}, in an attempt to reduce false positive findings. 
In \cref{appendix:correction_level}, we repeat the experiment of this section with $C = 2^6$ and find almost identical results. We hypothesize, that the effects of a reduced $C$ is more pronounced at larger $d$.
When $d = 100$, we test hypotheses with the correction factor $C(1)$ of \cref{theorem:restricted_search}.
In both cases, we test the hypothesis of invariance of the empty set at level $\alpha_0 = 10^{-6}$ (cf.\ \cref{remark:alpha0}). In \cref{appendix:alpha0}, we investigate the effects on the quantities $\mathbb{P}(\hat{S}_{\AS} \subseteq \AN_Y)$ and $\mathbb{P}(\hat{S}_{\AS}^1 \subseteq \AN_Y)$ when varying $\alpha_0$, confirming that choosing $\alpha_0$ too high can lead to a reduced probability of $\hat{S}_{\AS}$ being a subset of ancestors.

\begin{figure}[t]
    \centerline{%
    \includegraphics[width=.7\linewidth]{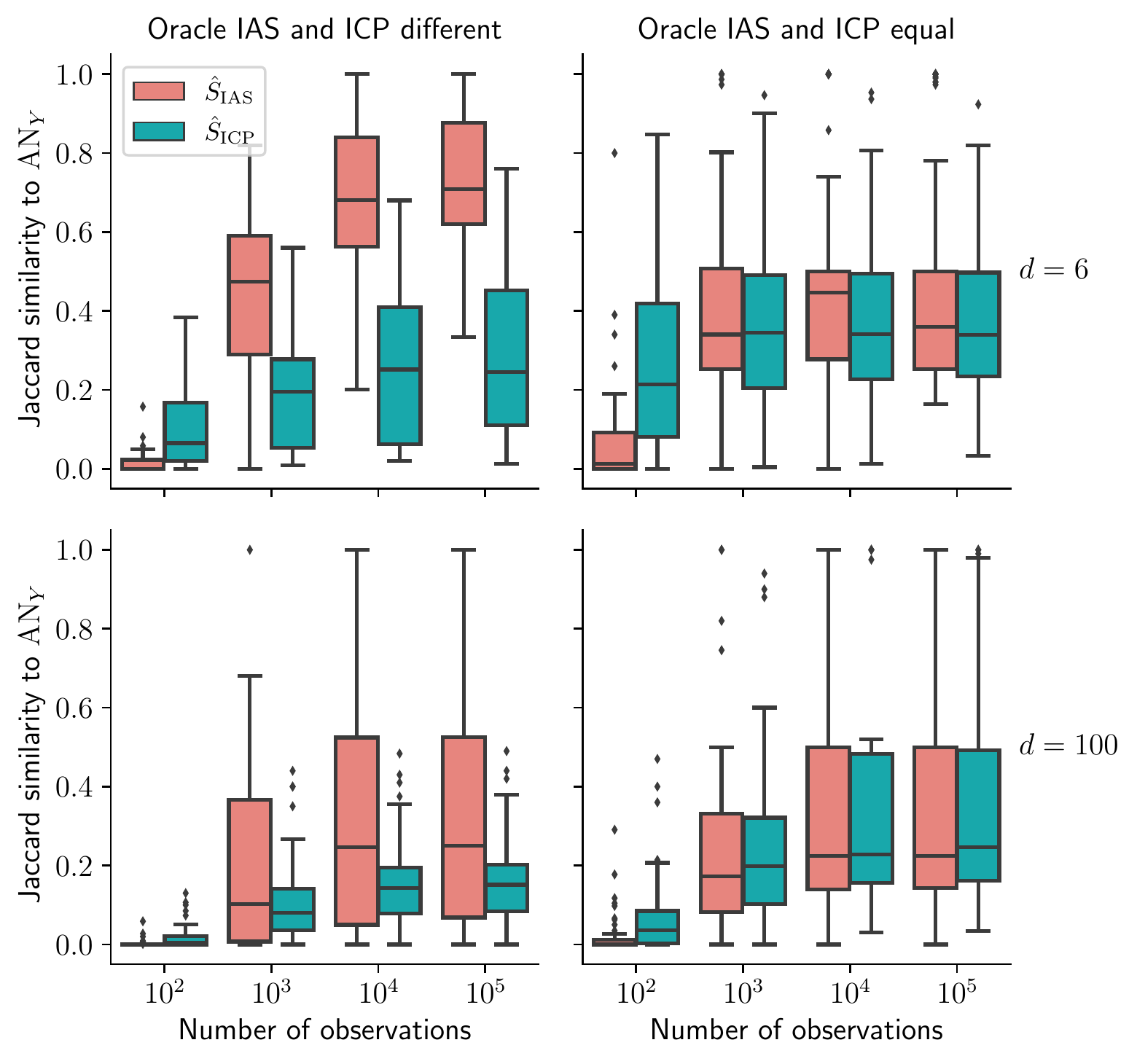}}
    \caption{%
    Comparison between the finite sample output of IAS and ICP and $\AN_Y$ on simulated data, see \cref{sec:experiment2}. 
    The plots show the Jaccard similarities between $\AN_Y$ and either $\hat{S}_{\AS}$ ($\hat{S}_{\AS}^1$ when $d = 100$) in red or $\hat{S}_{\ICP}$ ($\hat{S}_{\ICP}^{\hat{\MB}}$ when $d = 100$) in blue and $\AN_Y$. 
    When $S_{\ICP} \neq S_{\AS}$ (left column), $\hat{S}_{\AS}$ is more similar to $\AN_Y$ than $\hat{S}_{\ICP}$. The procedures are roughly equally similar to $\AN_Y$ when $S_{\ICP} = S_{\AS}$ (right column).
    Graphs represented in each boxplot: $42$ (top left), $58$ (top right), $40$ (bottom left) and $60$ (bottom right).}\label{fig:experiment2_jaccard}
\end{figure}

In \cref{fig:experiment2_jaccard} the results of the simulations are displayed. In SCMs where the oracle versions $S_{\AS}$ and $S_{\ICP}$ are not equal,
$\hat{S}_{\AS}$ achieved, on average, a higher Jaccard similarity to $\AN_Y$ than $\hat{S}_{\ICP}$.
This effect is less pronounced when $d = 100$. 
We believe that the difference in Jaccard similarities is more pronounced when using larger values of $m$.
When $S_{\AS} = S_{\ICP}$, the two procedures achieve roughly the same Jaccard similarities to $\AN_Y$, as expected. When the number of observations is one hundred, IAS generally fails to find any ancestors and outputs the empty set (see \cref{fig:appendix_exp2_main}), indicating that the we do not have power to reject the empty set when there are few observations. This is partly by design; we test the empty set for invariance at reduced level $\alpha_0$ in order to protect against making false positive findings when the environment has a weak effect on $Y$. However, even without testing the empty set at a reduced level, IAS has to correct for making multiple comparisons, contrary to ICP, thus lowering the marginal significance level each set is tested at. When computing the jaccard similarities with either $\alpha_0 = \alpha$ or $\alpha_0 = 10^{-12}$, the results were similar (not shown). 
We repeated the experiments with $d = 6$ with a weaker influence of the environment (do-interventions of strength $0.5$ instead of $1$) and found comparable results, with slightly less power in that the empty set is found more often,
see \cref{appendix:weakInstrument}.

We compare our method with a variant, called {\IASest}, where we first estimate (e.g., using methods proposed by \citealt{JCI} or \citealt{squires2020permutation}) a member graph of the Markov equivalence class (`I-MEC') and apply the oracle algorithm from \cref{sec:oracle-algorithm} (by reading of d-separations in that graph) to estimate $\MIP$.
In general, however, such an approach comes with additional assumptions; furthermore, even in the linear setup considered here, its empirical performance for large graphs is worse than the proposed method IAS, see \cref{app:UT-IGSP}.

\subsection{IAS in High Dimensional Genetic Data}\label{sec:experiment_gene}
\begin{figure}[t]
    \centering
    \includegraphics[width=0.7\linewidth]{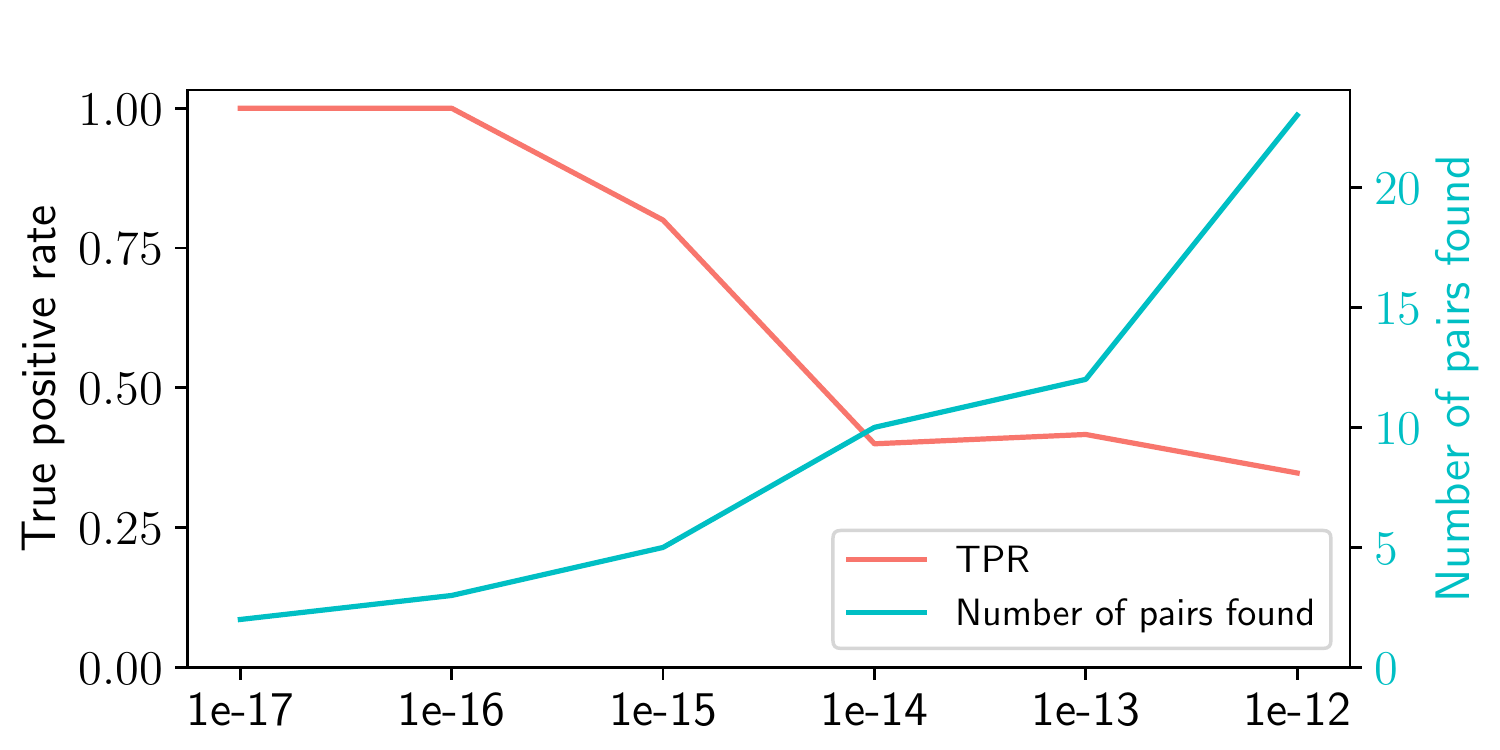}
    \caption{True positive rates and number of gene pairs found in the experiment in \cref{sec:experiment_gene}. 
    On the $x$-axis, we change $\alpha_0$, the threshold for invariance of the empty set. When $\alpha_0$ is small, we only search for pairs if the environment has a very significant effect on $Y$.
    For smaller $\alpha_0$, fewer pairs are found to be invariant (blue line), but those found, are more likely to be true positives (red line). This supports the claim that the lower $\alpha_0$ is, the more conservative our approach is. 
    }
    \label{fig:experiment-gene}
\end{figure}
We evaluate our approach in a data set on gene expression in yeast \cite{kemmeren2014large}. The data contain full-genome mRNA expressions of $d = 6{,}170$ genes and consists of $n_{\textrm{obs}} = 160$ unperturbed observations ($E = 0$) and $n_{\textrm{int}} = 1{,}479$ intervened-upon observations ($E = 1$); each of the latter observations correspond to the deletion of a single (known) gene. 
For each response
gene $\gene_Y \in [d]$, we apply 
the procedure from \cref{sec:higher-dimensions} with $m = 1$ to search for ancestors.

We first test for invariance of the empty set, i.e., whether the distribution of $\gene_Y$ differs between the observational and interventional environment. We test this at a conservative level $\alpha_0 = 10^{-12}$ in order to protect against a high false positive rate (see Remark~\ref{remark:alpha0}).
For $3{,}631$ out of $6{,}170$ response genes, the empty set is invariant, and we disregard them as response genes.

For each response gene, for which the empty set is not invariant, we 
apply our procedure. 
More specifically, when testing whether $\gene_X$ is an ancestor of $\gene_Y$, we exclude any observation in which either $\gene_X$ or $\gene_Y$ was intervened on.
We then test whether the empty set is still rejected, at level $\alpha_0 = 10^{-12}$,
and whether $\gene_X$ is invariant at level $\alpha = 0.25$. 
Since a set $\{\gene_X\}$ is deemed minimally invariant if the $p$-value exceeds $\alpha$, setting $\alpha$ large is conservative for the task of finding ancestors.
Indeed, 
when estimating $\hat{S}_{\AS}^m$, one can test the sets of size $m$ at a higher level $\alpha_1 > \alpha$. This is conservative, because falsely rejecting a minimally invariant set of size $m$ does not break the inclusion $\hat{S}_{\AS}^m \subseteq \AN_Y$. However, if one has little power against the non-invariant sets of size $m$, testing at level $\alpha_1$ can protect against false positives.\footnote{Only sets of size exactly $m$ can be tested at level $\alpha_1$; the remaining hypotheses should still be corrected by $C(m)$ (or by the hypothesized number of minimally invariant sets).}

We use the held-out data point, where $\gene_X$ is intervened on, to determine as ground truth, whether $\gene_X$ is indeed an ancestor of $\gene_Y$. We define $\gene_X$ as a true ancestor of $\gene_Y$ if the value of $\gene_Y$ when $\gene_X$ is intervened on, lies in the $q_{TP} = 1\%$ tails of the observational distribution of $\gene_Y$.

We find $23$ invariant pairs $(\gene_X, \gene_Y)$; of these, $7$ are true positives. 
In comparison, \citet{peters2016causal} applies ICP to the same data, and with the same definition of true positives. They predict $8$ pairs, of which $6$ are true positives. This difference is in coherence with the motivation put forward in \cref{sec:AS}: Our approach predicts many more ancestral pairs ($8$ for ICP compared to $23$ for IAS). Since ICP does not depend on power of the test, they have a lower false positive rate ($25\%$ for ICP compared to $69.6\%$ for IAS). 

In \cref{fig:experiment-gene}, we explore how changing $\alpha_0$ and $q_{TP}$
impacts the true positive rate. 
Reducing $\alpha_0$ increases the true positive rate, but lowers the number of gene pairs found (see \cref{fig:experiment-gene}). This is because a lower $\alpha_0$ makes it more difficult to detect non-invariance of the empty set, making the procedure more conservative (with respect to finding ancestors); see \cref{remark:alpha0}.
For example, when $\alpha_0 \leq 10^{-15}$, 
the true positive rate is above $0.8$; however, $5$ or fewer pairs are found. 
When searching for ancestors, the effect of intervening may be reduced by noise from intermediary variables, so $q_{TB} = 1\%$ might be too strict; in \cref{appendix:q-tb}, we analyze the impact of increasing $q_{TB}$.

\section{Extensions}\label{sec:extensions}

\subsection{Latent variables}\label{sec:hiddens}
In \Cref{assumption:1}, we assume that all variables $X$ are observed and 
that there are no hidden variables $H$. 
Let us write
$X = X_O \, \dot\cup \, X_H$, where only  $X_O$ is observed 
and define $\IP \coloneqq \{S \subseteq X_O \mid S \text{ invariant}\}$.
We can then define 
\begin{equation*}
    S_{\AS, O} \coloneqq \bigcup_{S\subseteq X_O: H_{0, S}^{\MIP} \text{ true}} S
\end{equation*}
(again with the convention that a union over the empty set is the
empty set),
and have the following modification of \cref{prop:tian1998}.
\begin{proposition}\label{prop:unobserved}
It holds that $S_{\AS, O} \subseteq \AN_Y$.
\end{proposition}

All results in this paper remain correct in the presence of hidden variables, except for \cref{theorem:MIP_properties} and \cref{prop:smallMIP_properties} (iii-iv).\footnote{These results do not hold in the presence of hidden variables, because it is not guaranteed that an invariant set exists among $X_O$ (e.g., consider a graph where all observed variables share a common, unobserved confounder with $Y$). However, if at least one minimally invariant set exists among the observed variables, 
then all results stated in this paper hold.}
Thus, the union of the observed minimally invariant sets, $S_{\AS, O}$ is a subset of $\AN_Y$ and can be learned from data in the same way as if no latent variables were present.

\subsection{Non-exogenous environments}
Throughout this paper, we have assumed that the environment variable is exogenous (\cref{assumption:1}). However, all of the results stated in this paper, except for \cref{lemma:S_ICP}, also hold under the alternative assumption that $E$ is an ancestor of $Y$, but not necessarily exogenous. From the remaining results, only the proof of \cref{theorem:simpleDef} uses exogeneity of $E$, but here the result follows from \citet{tian1998finding}. In all other proofs, we account for both options. This extension also remains valid in the presence of hidden variables, using the same arguments as in \cref{sec:hiddens}.

\section{Conclusion and Future Work}
Invariant Ancestry Search (IAS) provides a framework for searching for causal ancestors of a response variable $Y$ through finding
minimally invariant sets of predictors by exploiting the existence of exogenous heterogeneity. 
The set $S_{\AS}$ is a subset of the ancestors of $Y$, a superset of $S_{\ICP}$ and, contrary to $S_{\ICP}$, invariant itself. Furthermore, the hierarchical structure of minimally invariant sets allows IAS to  search for causal ancestors only among subsets up to a predetermined size. 
This avoids exponential runtime and allows us to apply the algorithm to large systems. 
We have shown that, asymptotically, $S_{\AS}$ can be identified from data with high probability if we are provided with a test for invariance that has asymptotic level and power. We have validated our procedure both on simulated and real data. 
Our proposed framework would benefit from further research in the maximal number of minimally invariant sets among graphs of a fixed size, as this would provide larger finite sample 
power for identifying ancestors. Further it is of interest to establish finite sample guarantees or convergence rates for IAS, possibly by imposing additional assumptions on the class of SCMs. 
Finally, even though current implementations are fast, it is an open theoretical question whether computing $S_{\AS}$ in the oracle setting of Section~\ref{sec:oracle-algorithm} is NP-hard, see Appendix~\ref{appendix:minimal_review}.

\section*{Acknowledgements}
NT and JP were supported by a research grant (18968) from VILLUM FONDEN.

\appendix
\onecolumn

\section{Proofs}\label{appendix:proofs}
\subsection{A direct Proof of \cref{theorem:simpleDef}}
\begin{proof}
Assume that $E$ is exogenous.
If $E \in \PA_Y$, then there are no minimally invariant sets, and the statement holds trivially. If $E \not\in \PA_Y$, then assume
for contradiction, that an invariant set $S_0\subsetneq S$ exists. By assumption, $|S\setminus S_0| > 1$, because otherwise $S_0$ would be non-invariant. 

We can choose $S_1 \subseteq S$ and $k_0, k_1, \ldots, k_l \in S$ with $l \geq 1$ such that for all $i = 1, \ldots, l: k_i \notin \DE_{k_0}$ and
\begin{align*}
    &S_0 \cup S_1 \cup \{k_0, \ldots, k_l\} = S &\in \IP \phantom{.}\\
    \text{for } 0 \leq i < l:\quad  &S_0 \cup S_1 \cup \{k_0, \ldots, k_i \} &\notin \IP \phantom{.} \\
    &S_0 \cup S_1 &\in \IP.
\end{align*}
This can be done by iteratively removing elements from $S\setminus S_0$, removing first the earliest elements in the causal order. The first invariant set reached in this process is then $S_0 \cup S_1$.

Since $S_0 \cup S_1 \cup \{k_0\}$ is non-invariant, there exists a path $\pi$ between $E$ and $Y$ that is open given $S_0 \cup S_1 \cup \{k_0\}$ but blocked given $S_0 \cup S_1$. Since removing $k_0$ blocks $\pi$, $k_0$ must be a collider or a descendant of a collider $c$ on $\pi$:
\begin{figure}[h]
    \centering
    \begin{tikzpicture}
        \node (E) at (0, 0) {$E$};
        \node (dots1) at (1, 0) {$\cdots$};
        \node (c) at (2, 0) {$c$};
        \node (dots2) at (3, 0) {$\cdots$};
        \node (Y) at (4, 0) {$Y$};
        \node (dots3) at (2, -0.8) {$\vdots$};
        \node (k0) at (2, -1.8) {$k_0$};
        \draw[->] (E) to (dots1);
        \draw[->] (dots1) to (c);
        \draw[->] (c) to (dots3);
        \draw[->] (dots3) to (k0);
        \draw[<-] (c) to (dots2);
        \draw[-] (dots2) to (Y);
        \draw [decorate,decoration={brace,amplitude=5pt,raise=2.5ex}] (0,0) -- (4,0) node[midway,yshift=2em]{$\pi$};
        \draw [decorate,decoration={brace,amplitude=5pt,mirror,raise=1ex}] (0,0) -- (2,0) node[midway,yshift=-2em]{$\pi_E$};
        \draw [decorate,decoration={brace,amplitude=5pt,mirror,raise=1ex}] (2,0) -- (4,0) node[midway,yshift=-2em]{$\pi_Y$};
    \end{tikzpicture}
\end{figure}

Here, $-$ represents an edge that either points left or right.
Since $\pi$ is open given $S_0\cup S_1$, the two sub-paths $\pi_E$ and $\pi_Y$ are open given $S_0\cup S_1$.

Additionally, since $S_0\cup S_1 \cup \{k_1, \ldots, k_l\} = S\setminus \{k_0\}$ is non-invariant, there exists a path $\tau$ between $E$ and $Y$ that is unblocked given $S_0\cup S_1 \cup \{k_1, \ldots, k_l\}$ and blocked given $S_0\cup S_1 \cup \{k_1, \ldots, k_l\}\cup \{k_0\}$. It follows that $k_0$ lies on $\tau$ (otherwise $\tau$ cannot be blocked by adding $k_0$) and $k_0$ has at least one outgoing edge. Assume, without loss of generality that there is an outgoing edge towards $Y$. 
\begin{figure}[h]
    \centering
    \begin{tikzpicture}
        \node (E) at (0, 0) {$E$};
        \node (dots1) at (1, 0) {$\cdots$};
        \node (k0) at (2, 0) {$k_0$};
        \node (dots2) at (3, 0) {$\cdots$};
        \node (Y) at (4, 0) {$Y$};
        \draw[->] (E) to (dots1);
        \draw[-] (dots1) to (k0);
        \draw[->] (k0) to (dots2);
        \draw[-] (dots2) to (Y);
        \draw [decorate,decoration={brace,amplitude=5pt,raise=2.5ex}] (0,0) -- (4,0) node[midway,yshift=2em]{$\tau$};
        \draw [decorate,decoration={brace,amplitude=5pt,mirror,raise=1ex}] (2,0) -- (4,0) node[midway,yshift=-2em]{$\tau_Y$};
    \end{tikzpicture}
\end{figure}
Since $\tau$ is open given $S_0\cup S_1 \cup \{k_1, \ldots, k_l\}$, so is $\tau_Y$. 

If there are no colliders on $\tau_Y$, then $\tau_Y$ is also open given $S_0\cup S_1$. But then the path the path $E \stackrel{\pi_E}{\cdots} \rightarrow c \rightarrow \cdots \rightarrow k_0 \stackrel{\tau_Y}{\rightarrow \cdots}$ is also open given $S_0 \cup S_1$, contradicting invariance of $S_0 \cup S_1$.

\begin{figure}[h]
    \centering
    \begin{tikzpicture}
        \node (E) at (0, 0) {$E$};
        \node (dots1) at (1, 0) {$\cdots$};
        \node (c) at (2, 0) {$c$};
        \node (dots2) at (3, -1.8) {$\cdots$};
        \node (Y) at (4, -1.8) {$Y$};
        \node (dots3) at (2, -0.8) {$\vdots$};
        \node (k0) at (2, -1.8) {$k_0$};
        \draw[->] (E) to (dots1);
        \draw[->] (dots1) to (c);
        \draw[->] (c) to (dots3);
        \draw[->] (dots3) to (k0);
        \draw[->] (k0) to (dots2);
        \draw[-] (dots2) to (Y);
        \draw [decorate,decoration={brace,amplitude=5pt,mirror,raise=1ex}] (0,0) -- (2,0) node[midway,yshift=-2em]{$\pi_E$};
        \draw [decorate,decoration={brace,amplitude=5pt,raise=1ex}] (2,-1.8) -- (4,-1.8) node[midway,yshift=2em]{$\tau_Y$};
    \end{tikzpicture}
\end{figure}

If there are colliders on $\tau_Y$, let $m$ be the collider closest to $k_0$, meaning that $m \in \DE_{k_0}$. Since $\tau_Y$ is open given $S_0 \cup S_1 \cup \{k_1, \ldots, k_l\}$, it means that either $m$ or a descendant of $m$ is in $S_0\cup S_1 \cup \{k_1, \ldots, k_l\}$. Since $\{k_1, \ldots, k_l\} \cap \DE_{k_0} = \emptyset$, there exist $v \in (S_0\cup S_1) \cap (\{m\} \cup \DE_m)$. But then $v \in \DE_{k_0} \cap (S_0 \cup S_1)$, meaning that $\pi$ is open given $S_0 \cup S_1$, contradicting invariance of $S_0 \cup S_1$.

We could assume that $\tau_Y$ had an outgoing edge from $k_0$ without loss of generality, because if there was instead an outgoing edge from $k_0$ on $\tau_E$, the above argument would work with $\pi_Y$ and $\tau_E$ instead. This concludes the proof.
\end{proof}

\subsection{A direct  proof of \cref{prop:tian1998}}
\begin{proof}

If $E$ is a parent of $Y$, we have
$\MIP = \emptyset$ and the statement follows trivially. Thus, assume that 
$E$ is not a parent of $Y$.
We will show that if $S\in \IP$ is not a subset of $\AN_Y$, then $S^* \coloneqq S \cap \AN_Y \in \IP$, meaning that $S \notin \MIP$.

Assume for contradiction that there is a path $p$ between $E$ and $Y$ that is 
open given $S^\ast$. 
Since $S \in \IP$, $p$ is blocked given $S$.
Then there exists a non-collider $Z$ on $p$ that is in $S \setminus \AN_Y$. We now argue that all nodes on $p$ are ancestors of $Y$, yielding a contradiction.

First, assume that there are no colliders on $p$. If $E$ is exogenous, then $p$ is directed from $E$ to $Y$. (If $E$ is an ancestor of $Y$, any node on $p$ is either an ancestor of $Y$ or $E$, and thus $Y$.) 
Second, assume that there are colliders 
on $p$. Since $p$ is open given the smaller set $S^\ast \subsetneq S$, 
all colliders on $p$ are in $S^*$ or have a descendant in $S^*$; therefore all colliders are ancestors of $Y$. 
If $E$ is exogenous, 
any node on $p$ is either an ancestor of $Y$ or of a collider on $p$. 
(If $E$ is an ancestor of $Y$, 
any node on $p$ is either an ancestor of $Y$, of a collider on $p$ or of $E$, and thus also $Y$.) 
This completes the proof of Proposition~\ref{prop:tian1998}.

\end{proof}

\subsection{Proof of \cref{theorem:MIP_properties}}\label{sec:proof_MIP_properties}
\begin{proof}
First, we show that $S_{\AS} \in \IP$. If $S_{\AS}$ is the union of a single minimally invariant set, it trivially holds that $S_{\AS}\in \IP$. Now assume that $S_{\AS}$ is the union of at least two minimally invariant sets, $S_{\AS} = S_1 \cup \ldots \cup S_n$, $n\geq 2$, and assume for a contradiction that there exists a path $\pi$ between $E$ and $Y$ that is unblocked given $S_{\AS}$. 

Since $\pi$ is blocked by a strict subset of $S_{\AS}$, it follows that $\pi$ has at least one collider; further every collider of $\pi$ is either in $S_{\AS}$ or has a descendant in $S_{\AS}$, and hence every collider of $\pi$ is an ancestor of $Y$, by \cref{prop:tian1998}. 
If $E$ is exogenous, 
$\pi$ has the following shape
\begin{figure}[h]
    \centering
    \begin{tikzpicture}
        \node (E) at (0, 0) {$E$};
        \node (dots1) at (1, 0) {$\cdots$};
        \node (c1) at (2, 0) {$c_1$};
        \node (dots2) at (3, 0) {$\cdots$};
        \node (c2) at (4, 0) {$c_2$};
        \node (dots3) at (5.5, 0) {$\cdots$};
        \node (c3) at (7,0) {$c_k$};
        \node (dots4) at (8,0) {$\cdots$};
        \node (Y) at (9,0) {$Y.$};
        
        \draw[->] (E) to (dots1);
        \draw[->] (dots1) to (c1);
        \draw[<-] (c1) to (dots2);
        \draw[->] (dots2) to (c2);
        \draw[<-] (c2) to (dots3);
        \draw[->] (dots3) to (c3);
        \draw[<-] (c3) to (dots4);
        \draw[->] (dots4) to (Y);
        
        \draw [decorate,decoration={brace,amplitude=5pt,raise=2.5ex}] (E) -- (c1) node[midway,yshift=2em]{$\pi_1$};
        \draw [decorate,decoration={brace,amplitude=5pt,raise=2.5ex}] (c1) -- (c2) node[midway,yshift=2em]{$\pi_2$};
        \draw [decorate,decoration={brace,amplitude=5pt,raise=2.5ex}] (c2) -- (c3) node[midway,yshift=2em]{$\pi_3, \ldots, \pi_k$};
        \draw [decorate,decoration={brace,amplitude=5pt,raise=2.5ex}] (c3) -- (Y) node[midway,yshift=2em]{$\pi_{k+1}$};
    \end{tikzpicture}
\end{figure}

(If $E$ is not exogenous but $E \in \AN_Y$, %
then $\pi$ takes either the form displayed above or
the shape displayed below.
\begin{figure}[h]
    \centering
    \begin{tikzpicture}
        \node (E) at (0, 0) {$E$};
        \node (dots1) at (1, 0) {$\cdots$};
        \node (c1) at (2, 0) {$c_1$};
        \node (dots2) at (3, 0) {$\cdots$};
        \node (c2) at (4, 0) {$c_2$};
        \node (dots3) at (5.5, 0) {$\cdots$};
        \node (c3) at (7,0) {$c_k$};
        \node (dots4) at (8,0) {$\cdots$};
        \node (Y) at (9,0) {$Y$.};
        
        \draw[<-] (E) to (dots1);
        \draw[->] (dots1) to (c1);
        \draw[<-] (c1) to (dots2);
        \draw[->] (dots2) to (c2);
        \draw[<-] (c2) to (dots3);
        \draw[->] (dots3) to (c3);
        \draw[<-] (c3) to (dots4);
        \draw[->] (dots4) to (Y);
        
        \draw [decorate,decoration={brace,amplitude=5pt,raise=2.5ex}] (E) -- (c1) node[midway,yshift=2em]{$\pi_1$};
        \draw [decorate,decoration={brace,amplitude=5pt,raise=2.5ex}] (c1) -- (c2) node[midway,yshift=2em]{$\pi_2$};
        \draw [decorate,decoration={brace,amplitude=5pt,raise=2.5ex}] (c2) -- (c3) node[midway,yshift=2em]{$\pi_3, \ldots, \pi_k$};
        \draw [decorate,decoration={brace,amplitude=5pt,raise=2.5ex}] (c3) -- (Y) node[midway,yshift=2em]{$\pi_{k+1}$};
    \end{tikzpicture}
\end{figure}
However, no matter which of the shapes $\pi$ takes, the proof proceeds the same.)
The paths $\pi_1, \ldots, \pi_{k+1}$, $k\geq 1$, do not have any colliders and are unblocked given $S_{\AS}$. In particular, $\pi_1, \ldots, \pi_{k+1}$ are unblocked given $S_1$.

The path $\pi_{k+1}$ must have a final edge pointing to $Y$, because otherwise it would be a directed path from $Y$ to $c_k$, which contradicts acyclicity since $c_k$ is an ancestor of $Y$.

As $c_1$ is an ancestor of $Y$, there exists a directed path, say $\rho_1$, from $c_1$ to $Y$. Since $\pi_1$ is open given $S_1$ and since $S_1$ is invariant, it follows that $\rho_1$ must be blocked by $S_1$ (otherwise the path $E \stackrel{\pi_1}{\rightarrow} c_1 \stackrel{\rho_1}{\rightarrow} Y$ would be open). For this reason, $S_1$ contains a descendant of the collider $c_1$. 

Similarly, if $\rho_2$ is a directed path from $c_2$ to $Y$, then $S_1$ blocks $\rho_2$, because otherwise the path $E \stackrel{\pi_1}{\rightarrow} c_1 \stackrel{\pi_2}{\leftarrow \cdots \rightarrow} c_2 \stackrel{\rho_2}{\rightarrow} Y$ would be open. Again, for this reason, $S_1$ contains a descendant of $c_2$. 

Iterating this argument, it follows that $S_1$ contains a descendant of every collider on $\pi$, and since $\pi_1, \ldots, \pi_{k+1}$ are unblocked by $S_1$, $\pi$ is open given $S_1$. This contradicts invariance of $S_1$ and proves that $S_{\AS} \in \IP$. 

We now show that $S_{\ICP} \subseteq S_{\AS}$ with equality if and only if $S_{\ICP} \in \IP$. First, $S_{\ICP} \subseteq S_{\AS}$ because $S_{\AS}$ is a union of the minimally invariant sets, and $S_{\ICP}$ is the intersection over all invariant sets. We now show the equivalence statement. 

Assume first that $S_{\ICP} \in \IP$. As $S_{\ICP}$ is the intersection of all invariant sets, $S_{\ICP} \in \IP$ implies that there exists exactly one invariant set, that is contained in all other invariant sets. By definition, this means that there is only one minimally invariant set, and that this set is exactly $S_{\ICP}$. Thus, $S_{\AS} = S_{\ICP}$.

Conversely assume that $S_{\ICP} \notin \IP$. By construction, $S_{\ICP}$ is contained in any invariant set, in particular in the minimally invariant sets. However, since $S_{\ICP}$ is not invariant itself, this containment is strict, and it follows that $S_{\ICP} \subsetneq S_{\AS}$.

\end{proof}

\subsection{Proof of \cref{lemma:S_ICP}}
\begin{proof}
First we show $\PA_Y \cap \left(\CH_E \cup \PA(\AN_Y\cap\CH_E)\right) \subseteq S_{\ICP}$. If $j \in \PA_Y\cap \CH_E$, any invariant set contains $j$, because otherwise the path $E \rightarrow j \rightarrow Y$ is open. Similarly, if $j \in \PA_Y \cap \PA(\AN_Y \cap \CH_E)$, any invariant set contains $j$ (there exists a node $j'$ such that $E \rightarrow j' \rightarrow \cdots \rightarrow Y$ and $E \rightarrow j' \leftarrow j \rightarrow Y$, 
and any invariant set $S$ must contain $j'$ or one of its descendants; thus, it must also contain $j$ to ensure that the path $E \rightarrow j' \leftarrow j \rightarrow Y$ is blocked by $S$.)
It follows that for all invariant $S$,
\begin{align*}
    \PA_Y \cap \left(\CH_E \cup \PA(\AN_Y\cap\CH_E)\right)\subseteq S,
\end{align*}
such that
\begin{align*}
    \PA_Y \cap \left(\CH_E \cup \PA(\AN_Y\cap\CH_E)\right)\subseteq \bigcap_{S \text{ invariant}}S.
\end{align*}
To show $S_{\ICP} \subseteq \PA_Y \cap \left(\CH_E \cup \PA(\AN_Y\cap\CH_E)\right)$, take any $j\notin \PA_Y \cap \left(\CH_E \cup \PA(\AN_Y\cap\CH_E)\right)$. We argue, that an invariant set $\bar{S}$ not containing $j$ exists, such that $j \notin S_{\ICP}=\bigcap_{S \text{ invariant}}S$. If $j \notin\PA_Y$, let $\bar{S}=\PA_Y$, which is invariant. 
If $j \in \PA_Y$, define
\begin{equation*}
    \bar{S} = (\PA_Y\setminus\{j\}) \cup \PA_j \cup (\CH_j \cap \AN_Y) \cup \PA(\CH_j \cap \AN_Y).
\end{equation*}
Because $j \notin \CH_E$ and $j\notin \PA(\AN_Y\cap\CH_E)$, we have $E \notin \bar{S}$. Also observe that $\bar{S}\subseteq \AN_Y$.
We show that any path between $E$ and $Y$ is blocked by $\bar{S}$, by considering all possible paths:
\begin{description}
    \item[$\cdots\mathbf{j'\rightarrow Y}$ for $\mathbf{j'\neq j}$:] Blocked because $j'\in \PA_Y\setminus\{j\}$.
    \item[$\mathbf{\cdots v\rightarrow j\rightarrow Y}$:] Blocked because $v\in\PA_j\subseteq \bar{S}$ and $E\notin \PA_j$.
    \item[$\mathbf{\cdots v\rightarrow c \leftarrow j\rightarrow Y}$ and $\mathbf{c\in\AN_Y}$:] Blocked because $v\in\PA_j(\CH_j\cap\AN_Y)$.
    \item[$\mathbf{\cdots v\rightarrow c \leftarrow j\rightarrow Y}$ and $\mathbf{c\notin\AN_Y}$:] Blocked because $\bar{S}\subseteq\AN_Y$, and since $c\notin\AN_Y$, $\bar{S}\cap\DE_c=\emptyset$ and the path is blocked given $\bar{S}$ because of the collider $c$.
    \item[$\mathbf{\cdots \rightarrow c \leftarrow \cdots \leftarrow v \leftarrow j\rightarrow Y}$ and $\mathbf{c\in\AN_Y}$:] Blocked because $v\in\AN_c$ and $c\in\AN_Y$, so $v\in\CH_j\cap\AN_Y\subseteq\bar{S}$.
    \item[$\mathbf{\cdots \rightarrow c \leftarrow \cdots \leftarrow v \leftarrow j\rightarrow Y}$ and $\mathbf{c\notin\AN_Y}$:] Same reason as for the case `$\mathbf{\cdots v\rightarrow c \leftarrow j\rightarrow Y}$ and $\mathbf{c\notin\AN_Y}$'.
    \item[$\mathbf{\cdots \rightarrow c \leftarrow \cdots \leftarrow Y}$] Since $\bar{S}\subseteq \AN_Y$, we must have $\bar{S}\cap\DE_c = \emptyset$ (otherwise this would create a directed cycle from $Y \rightarrow \cdots \rightarrow Y$). Hence the path is blocked given $\bar{S}$ because of the collider $c$.
\end{description}
Since there are no open paths from $E$ to $Y$ given $\bar{S}$, $\bar{S}$ is invariant, and $S_{\ICP}\subseteq \bar{S}$. Since $j \notin \bar{S}$, it follows that $j \notin S_{\ICP}$. This concludes the proof.
\end{proof}

\subsection{Proof of \cref{theorem:MIP_test}}
\begin{proof}
Consider first the case where all marginal tests have pointwise asymptotic power and pointwise asymptotic level.

\textbf{Pointwise asymptotic level:}
Let $\mathbb{P}_0 \in H_{0, S}^{\MIP}$. By the assumption of pointwise asymptotic level, there exists a non-negative sequence $(\epsilon_n)_{n \in \mathbb{N}}$ such that $\lim_{n \to \infty} \epsilon_n = 0$ and $\mathbb{P}_0(\phi_n(S) = 1) \leq \alpha + \epsilon_n$.
Then
    \begin{align*}
        \mathbb{P}_0
        (\phi_{n}^{\MIP}(S) = 1) &= 
            \mathbb{P}_0
            \left(
            (\phi_n(S) = 1)
    \cup
    \bigcup_{j \in S}
    (\phi_n(S \setminus \{j\}) = 0)
    \right) \\
     & \leq \mathbb{P}_0
     (\phi_n(S) = 1) +
        \sum_{j \in S} \mathbb{P}_0
        (\phi_n(S \setminus \{j\}) = 0) \\
    & \leq \alpha + \epsilon_n +
        \sum_{j \in S} \mathbb{P}_0
        (\phi_n(S \setminus \{j\}) = 0) \\
    & \rightarrow \alpha + 0 \qquad \text{as} \,\, n \to \infty\\
    & = \alpha.
    \end{align*}
The convergence step follows from
    \begin{equation*}
    H_{0, S}^{\MIP} = H_{0, S}^{\IP} \cap \bigcap_{j \in S} H_{A, S\setminus \{j\}}^{\IP}
    \end{equation*}
and from the assumption of pointwise asymptotic level and power. As $\mathbb{P}_0 \in H_{0, S}^{\MIP}$ was arbitrary, this shows that $\phi_n^{\MIP}$ has pointwise asymptotic level.

\textbf{Pointwise asymptotic power:}
To show that the decision rule has pointwise asymptotic power, consider any $\mathbb{P}_A \in H_{A,S}^{\MIP}$. We have that
    \begin{equation}\label{eq:altsplitup}
        H_{A, S}^{\MIP} =
        H_{A, S}^{\IP} \cup \left(
        H_{0, S}^{\IP} \cap \bigcup_{j \in S} H_{0, S \setminus \{j\}}^{\IP}
        \right).
    \end{equation}
As the two sets $H_{A, S}^{\IP}$
and
    \begin{equation*}
        H_{0, S}^{\IP} \cap \bigcup_{j \in S} H_{0, S\setminus \{j\}}^{\IP}
    \end{equation*}
are disjoint, we can consider them one at a time. Consider first the case $\mathbb{P}_{A} \in H_{A, S}^{\IP}$. This means that $S$ is not invariant and thus
    \begin{align*}
        \mathbb{P}_{A} (\phi_{n}^{\MIP}(S) = 1)  &=
            \mathbb{P}_{A} \left(
            (\phi_n(S) = 1)
            \cup
            \bigcup_{j \in S}
            (\phi_n(S \setminus \{j\}, \alpha) = 0)
        \right) \\
        & \geq \mathbb{P}_{A}(\phi_n(S) = 1) \\ 
        & \to 1 \qquad \text{as} \,\, n \to \infty
    \end{align*}
by the assumption of pointwise asymptotic power. 

Next, assume that 
there exists $j' \in S$ such that 
$\mathbb{P}_{A} \in (H_{0, S}^{\IP} \cap H_{0, S \setminus \{j'\}}^{\IP})$. Then, %
    \begin{align*}
    \mathbb{P}_{A} (\phi_{n}^{\MIP}(S) = 1) &=
            \mathbb{P}_{0} \left(
            (\phi_n(S) = 1)
    \cup
    \bigcup_{j \in S}
    (\phi_n(S \setminus \{j\}) = 0)
    \right) \\ 
    & \geq \mathbb{P}_A(\phi_n(S \setminus \{j'\}) = 0) \\
    & \geq 1 - \alpha - \epsilon_n \\
    & \rightarrow 1 - \alpha \qquad \text{as} \,\, n \to \infty.
    \end{align*}
Thus, for arbitrary $\mathbb{P}_A \in H_{A, S}^{\MIP}$ we have shown that $\mathbb{P}_A(\phi_n^{\MIP}(S) = 1) \geq 1 - \alpha$ in the limit. This shows that $\phi_n^{\MIP}$ has pointwise asymptotic power of at least $1 - \alpha$.
This concludes the argument for pointwise asymptotic power. 

Next, consider the case that the marginal tests have uniform asymptotic power and uniform asymptotic level. The calculations for showing that $\phi_n^{\MIP}$ has uniform asymptotic level and uniform asymptotic power of at least $1 - \alpha$ are almost identical to the pointwise calculations.

\textbf{Uniform asymptotic level:}
By the assumption of uniform asymptotic level, there exists a non-negative sequence $\epsilon_n$ such that $\lim_{n \rightarrow \infty} \epsilon_n = 0$ and $\sup_{\mathbb{P} \in H_{0, S}^{\IP} } \mathbb{P}(\phi_n(S) = 1) \leq \alpha + \epsilon_n$. Then, 
\begin{align*}
    \sup\limits_{\mathbb{P} \in H_{0, S}^{\MIP}} \mathbb{P}
        (\phi_{n}^{\MIP}(S) = 1) &= 
            \sup\limits_{\mathbb{P} \in H_{0, S}^{\MIP}} \mathbb{P}
            \left(
            (\phi_n(S) = 1)
    \cup
    \bigcup_{j \in S}
    (\phi_n(S \setminus \{j\}) = 0)
    \right) \\
     & \leq \sup\limits_{\mathbb{P} \in H_{0, S}^{\MIP}} \left(\mathbb{P}
     (\phi_n(S) = 1) +
        \sum_{j \in S} \mathbb{P}
        (\phi_n(S \setminus \{j\}) = 0)\right) \\
    & \leq \sup\limits_{\mathbb{P} \in H_{0, S}^{\MIP}} \mathbb{P}
     (\phi_n(S) = 1) +
        \sum_{j \in S} \sup\limits_{\mathbb{P} \in H_{0, S}^{\MIP}}\mathbb{P}
        (\phi_n(S \setminus \{j\}) = 0) \\
    & \leq \alpha + \epsilon_n +
        \sum_{j \in S} \left(1 - \inf\limits_{\mathbb{P} \in H_{0, S}^{\MIP}}\mathbb{P}
        (\phi_n(S \setminus \{j\}) = 1)\right) \\
    & \rightarrow \alpha + 0 + \sum_{j \in S} (1 - 1) \qquad \text{as} \,\, n \to \infty\\
    & = \alpha.
    \end{align*}

\textbf{Uniform asymptotic power:}
From~\eqref{eq:altsplitup}, it follows that
    \begin{equation*}
        \inf\limits_{\mathbb{P} \in H_{A, S}^{\MIP}} \mathbb{P} (\phi_{n}^{\MIP}(S) = 1) =
        \min\left\{
        \inf\limits_{\mathbb{P} \in H_{A, S}^{\IP}} \mathbb{P} (\phi_{n}^{\MIP}(S) = 1),
        \inf\limits_{\mathbb{P} \in H_{0, S}^{\IP} \cap \bigcup_{j \in S} H_{0, S \setminus \{j\}}^{\IP}} \mathbb{P} (\phi_{n}^{\MIP}(S) = 1)
        \right\}.
    \end{equation*}
We consider the two inner terms in the above separately. First,
    \begin{align*}
        \inf\limits_{\mathbb{P} \in H_{A, S}^{\IP}} \mathbb{P} (\phi_{n}^{\MIP}(S) = 1)  &=
            \inf\limits_{\mathbb{P} \in H_{A, S}^{\IP}}
            \mathbb{P} \left(
            (\phi_n(S) = 1)
            \cup
            \bigcup_{j \in S}
            (\phi_n(S \setminus \{j\}) = 0)
        \right) \\
        & \geq \inf\limits_{\mathbb{P} \in H_{A, S}^{\IP}} \mathbb{P}(\phi_n(S) = 1) \\
        & \to 1  \qquad \text{as} \,\, n \to \infty.
    \end{align*}
Next, 
    \begin{align*}
       \inf\limits_{\mathbb{P} \in H_{0, S}^{\IP} \cap \bigcup_{j \in S} H_{0, S \setminus \{j\}}^{\IP}} \mathbb{P} (\phi_{n}^{\MIP}(S) = 1)  &=
       \inf\limits_{\mathbb{P} \in H_{0, S}^{\IP} \cap \bigcup_{j \in S} H_{0, S \setminus \{j\}}^{\IP}}
            \mathbb{P} \left(
            (\phi_n(S) = 1)
            \cup
            \bigcup_{j \in S}
            (\phi_n(S \setminus \{j\}) = 0)
        \right) \\
        & = \min\limits_{j \in S} \left\{
        \inf\limits_{\mathbb{P} \in H_{0, S}^{\IP} \cap H_{0, S \setminus \{j\}}^{\IP}} 
            \mathbb{P} \left(
            (\phi_n(S) = 1)
            \cup
            \bigcup_{j \in S}
            (\phi_n(S \setminus \{j\}) = 0)
        \right)
        \right\} \\
        & \geq \min\limits_{j \in S} \left\{
        \inf\limits_{\mathbb{P} \in H_{0, S}^{\IP} \cap H_{0, S \setminus \{j\}}^{\IP}} 
            \mathbb{P} (\phi_n(S \setminus \{j\}) = 0)
        \right\}
        \\
        & = \min\limits_{j \in S} \left\{
        1 - \sup\limits_{\mathbb{P} \in H_{0, S}^{\IP} \cap H_{0, S \setminus \{j\}}^{\IP}} 
        \mathbb{P}(\phi_n(S \setminus \{j\}) = 1)
        \right\} \\
        & \geq 1 - \alpha - \epsilon_n \\
        & \rightarrow 1 - \alpha \qquad \text{as} \,\, n \to \infty.
    \end{align*}
This shows that $\phi_n^{\MIP}$ has uniform asymptotic power of at least $1 - \alpha$, which completes the proof.
\end{proof}

\subsection{Proof of \cref{theorem:exhaustive_search}}
\begin{proof}
We have that
\begin{equation*}
    \lim\limits_{n \to \infty} \mathbb{P}(\hat{S}_{\AS} \subseteq \AN_Y) \geq
    \lim\limits_{n \to \infty} \mathbb{P}(\hat{S}_{\AS} = S_{\AS})
\end{equation*}
as $S_{\AS} \subseteq \AN_Y$ by \cref{theorem:MIP_properties}. Furthermore, we have 
    \begin{equation*}
        \mathbb{P}(\hat{S}_{\AS} = S_{\AS}) \geq \mathbb{P}(\widehat{\MIP} = \MIP).
    \end{equation*}
Let $A \coloneqq \{S \mid S \not\in \IP\} \setminus \{S \mid \exists S' \subsetneq S \text{ s.t.\ } S' \in \MIP \}$ be those non-invariant sets that do not contain a minimally invariant set and observe that
    \begin{equation}\label{eq:MIPevent0}
        (\widehat{\MIP} = \MIP) \supseteq \bigcap\limits_{S \in \MIP} (\phi_n(S, \alpha C^{-1}) = 0) \cap \bigcap\limits_{S \in A} (\phi_n(S, \alpha C^{-1}) = 1).
    \end{equation}
To see why this is true, note that to correctly recover $\MIP$, we need to 1) accept the hypothesis of minimal invariance for all minimally invariant sets and 2) reject the hypothesis of invariance for all non-invariant sets that are not supersets of a minimally invariant set (any superset of a set for which the hypothesis of minimal invariance is not rejected is removed in the computation of $\widehat{\MIP}$). 
Then,
    \begin{align*}
    \mathbb{P}(\widehat{\MIP} = \MIP) & \geq 
    \mathbb{P}\left(
        \bigcap\limits_{S \in \MIP} (\phi_n(S, \alpha C^{-1}) = 0)
        \cap
        \bigcap\limits_{S \in A} (\phi_n(S, \alpha C^{-1}) = 1)
    \right)
    \\
    & \geq 1 - \mathbb{P}
    \left(
        \bigcup\limits_{S \in \MIP} (\phi_n(S, \alpha C^{-1}) = 1)
    \right) - \sum\limits_{S \in A} \mathbb{P}(\phi_n(S, \alpha C^{-1}) = 0)
     \\
    & \geq 1 - \sum\limits_{S \in \MIP} \mathbb{P}(\phi_n(S, \alpha C^{-1}) = 1)
    - \sum\limits_{S \in A} \mathbb{P}(\phi_n(S, \alpha C^{-1}) = 0)
     \\
    & \geq 1 - \sum\limits_{S \in \MIP} (\alpha C^{-1} + \epsilon_{n, S})
    - \sum\limits_{S \in A} \mathbb{P}(\phi_n(S, \alpha C^{-1}) = 0)
     \\
    & \geq 1 - \vert\MIP\vert \alpha C^{-1} +  \sum\limits_{S \in \MIP}\epsilon_{n, S}
    - \sum\limits_{S \in A} \mathbb{P}(\phi_n(S, \alpha C^{-1}) = 0)
     \\
    & \geq 1 - \alpha + \sum\limits_{S \in \MIP}\epsilon_{n, S}
    - \sum\limits_{S \in A} \mathbb{P}(\phi_n(S, \alpha C^{-1}) = 0)
     \\
    & \rightarrow 1 - \alpha \quad \text{as } n \rightarrow \infty,
    \end{align*}
where $(\epsilon_{n, S})_{n \in \mathbb{N}, S \in \MIP}$ are non-negative sequences that converge to zero and the last step follows from the assumption of asymptotic power. The sequences $(\epsilon_{n, S})_{n \in \mathbb{N}, S \in \MIP}$ exist by the assumption of asymptotic level.
\end{proof}

\subsection{Proof of \cref{prop:smallMIP_properties}}
\begin{proof}
We prove the statements one by one.

\paragraph{(i)} Since $S_{\AS}^m$ is the union over some of the minimally invariant sets, $S_{\AS}^m \subseteq S_{\AS}$. Then the statement follows from \cref{prop:tian1998}.

\paragraph{(ii)} If $m \geq m_{\max}$, all $S\in\MIP$ satisfy the requirement $\vert S \vert \leq m$.

\paragraph{(iii)} If $m \geq m_{\min}$, then $S_{\AS}^m$ contains at least one minimally invariant set. The statement then follows from the first part of the proof of \cref{theorem:MIP_properties} given in \cref{sec:proof_MIP_properties}.

\paragraph{(iv)} $S_{\AS}^m$ contains at least one minimally invariant set and, by (iii), it is itself invariant. Thus, if $S_{\ICP}\not\in\IP$, then $S_{\ICP} \subsetneq S_{\AS}^m$. If $S_{\ICP} \in \IP$, then there exists only one minimally invariant set, which is $S_{\ICP}$ (see proof of \cref{theorem:MIP_properties}), and we have $S_{\ICP} = S_{\AS}^m$. This concludes the proof. 
\end{proof}

\subsection{Proof of \cref{theorem:restricted_search}}
\begin{proof}
The proof is identical to the proof of \cref{theorem:exhaustive_search}, when changing the correction factor $2^{-d}$ to $C(m)^{-1}$, adding superscript $m$'s to the quantities $\widehat{\MIP}$, $\hat{S}_{\AS}$ and $S_{\AS}$, and adding the condition $\vert S \vert \leq m$ to all unions, intersections and sums.
\end{proof}

\subsection{Proof of \cref{prop:unobserved}}
By \cref{prop:tian1998}, we have $S_{\AS} \subseteq \AN_Y$, and since $S_{\AS, O} \subseteq S_{\AS}$, the claim follows immediately.

\section{Oracle Algorithms for Learning $S_{\AS}$}\label{appendix:minimal_review}
In this section, we review some of the existing literature on minimal $d$-separators, which can be exploited to give an algorithmic approach for finding $S_{\AS}$ from a DAG. We first introduce the concept of $M$-minimal separation with respect to a constraining set $I$.
\begin{definition}[\citet{van2019separators}, Section 2.2]
Let 
$I \subseteq [d]$,
$K \subseteq [d]$, 
and $S \subseteq  [d]$. We say that $S$ is a \emph{$K$-minimal separator} of $E$ and $Y$ with respect to a constraining set $I$ if all of the following are true:
    \begin{compactitem}
        \item[(i)] $I \subseteq S$.
        \item[(ii)] $S \in \IP$.
        \item[(iii)] There does not exists $S' \in \IP$ such that $K \subseteq S' \subsetneq S$.
    \end{compactitem}
    We denote by $M_{K,I}$ the set of all $K$-minimal separating sets with respect to constraining set $I$.
\end{definition}
(In this work, $S \in \IP$ means $E \indep Y\,|\,S$, but it can stand for other separation statements, too.) 
The definition of a $K$-minimal separator coincides with the definition of a minimally invariant set if both $K$ and the constraining set $I$ are equal to the empty set.
An $\emptyset$-minimal separator with respect to constraining set $I$ is called a \emph{strongly-minimal separator with respect to constraining set $I$}.

We can now represent \eqref{eq:S_AS} using this notation.
$M_{\emptyset,\emptyset}$ contains the minimally invariant sets and thus
    \begin{equation*}
        S_{\AS} \coloneqq \bigcup\limits_{S \in M_{\emptyset,\emptyset}} S.
    \end{equation*}
Listing the set $M_{I,I}$ of all $I$-minimal separators with respect to the constraining set $I$ (for any $I$) can be done in polynomial delay time $\mathcal{O}(d^3)$  \citep{van2019separators, Takata2010}, where  delay here means that finding the next element of $M_{I,I}$ (or announcing that there is no further element) has cubic complexity. This is the algorithm we exploit, as described in the main part of the paper.

Furthermore, 
we have
$$
i \in S_{\AS} \quad \Leftrightarrow \quad
M_{\emptyset, \{i\}} \neq \emptyset.
$$ 
This is because $i \in S_{\AS}$ if and only if 
there is a minimally invariant set  that contains $i$, which is the case if and only if there exist a strongly minimal separating set with respect to constraining set $\{i\}$. 
Thus, we can construct $S_{\AS}$ 
by checking, for each $i$, whether 
there is an element in
$M_{\emptyset, \{i\}}$.
Finding a strongly-minimal separator with respect to constraining set $I$, 
i.e., 
finding an element in 
$M_{\emptyset, I}$, 
is NP-hard if the set $I$ is allowed to grow 
\citep{van2019separators}.
To the best of our knowledge, however, 
it is unknown whether 
finding an element in 
$M_{\emptyset, \{i\}}$, for a singleton $\{i\}$ is NP-hard.

\section{The Maximum Number of Minimally
Invariant Sets}\label{appendix:simulate_max}
If one does not have a priori knowledge about the graph of the system being analyzed, one can still apply \cref{theorem:exhaustive_search} with a correction factor $2^d$, as this ensures (with high probability) that no minimally invariant sets are falsely rejected. However, we know that the correction factor is strictly conservative, as there cannot exist $2^d$ minimally invariant sets in a graph. Thus, correcting for $2^d$ tests, controls the familywise error rate (FWER) among minimally invariant sets, but increases the risk of falsely accepting a non-invariant set relatively more than what is necessary to control the FWER. Here, we discuss the maximum number of minimally invariant sets that can exist in a graph with $d$ predictor nodes and how a priori knowledge about the sparsity of the graph and the number of interventions can be leveraged to estimate a less strict correction that still controls the FWER.

As minimally invariant sets only contain ancestors of $Y$ (see \cref{prop:tian1998}), we only need to consider graphs where $Y$ comes last in a causal ordering. Since $d$-separation is equivalent to undirected separation in the moralized ancestral graph \citep{lauritzen1996graphical}, finding the largest number of minimally invariant sets is equivalent to finding the maximum number of minimal separators in an undirected graph with $d + 2$ nodes. It is an open question how many minimal separators exists in a graph with $d + 2$ nodes, but it is known that a lower bound for the maximum number of minimal separators is in $\Omega(3^{d/3})$ \citep{gaspers2015number}. We therefore propose using a correction factor of $C = 3^{\lceil d/3 \rceil}$ when estimating the set $\hat{S}_{\AS}$ from \cref{theorem:exhaustive_search} if one does not have a priori knowledge of the number of minimally invariant sets in the DAG of the SCM being analyzed. 
This is a heuristic choice and is not conservative for all graphs.

\cref{theorem:exhaustive_search} assumes asymptotic power of the invariance test, but as we can only have a finite amount of data, we will usually not have full power against all non-invariant sets that are not supersets of a minimally invariant set. Therefore, choosing a correction factor that is potentially too low represents a trade-off between error types: if we correct too little, we stand the risk of falsely rejecting a minimally invariant set but not rejecting a superset of it, whereas when correcting too harshly, there is a risk of failing to reject non-invariant sets due to a lack of power.

If one has a priori knowledge of the sparsity or the number of interventions, these can be leveraged to estimate the maximum number of minimally invariant sets using simulation, by the following procedure:
\begin{enumerate}
    \item For $b = 1, \dots, B$:
    \begin{enumerate}
        \item Sample a DAG with $d$ predictor nodes, $N_{\textrm{interventions}} \sim \mathbb{P}_N$ interventions and $p \sim \mathbb{P}_p$ probability of an edge being present in the graph over $(X, Y)$, such that $Y$ is last in a causal ordering. The measures $\mathbb{P}_N$ and $\mathbb{P}_p$ are distributions representing a priori knowledge. For instance, in a controlled experiment, the researcher may have chosen the number $N_0$ of interventions. Then, $\mathbb{P}_N$ is a degenerate distribution with $\mathbb{P}_N(N_0)=1$.
        \item Compute the set of all minimally invariant sets, e.g., using the \verb|adjustmentSets| algorithm from \texttt{dagitty} \citep{dagitty}.
        \item Return the number of minimally invariant sets.
    \end{enumerate}
    \item Return the largest number of minimally sets found in the $B$ repetitions above.
\end{enumerate}
Instead of performing $B$ steps, one can  continually update the largest number of minimally invariant sets found so far and end the procedure if the maximum has not updated in a predetermined number of steps, for example.

\section{A Finite Sample Algorithm for Computing $\hat{S}_{\AS}$}\label{appendix:S_AS_algorithm}
In this section, we provide an algorithm for computing the sets $\hat{S}_{\AS}$ and $\hat{S}_{\AS}^m$ presented in \cref{theorem:exhaustive_search,theorem:restricted_search}. The algorithm finds minimally invariant sets by searching for invariant sets among sets of increasing size, starting from the empty set. This is done, because the first (correctly) accepted invariant is a minimally invariant set. Furthermore, any set that is a superset of an accepted invariant set, does not need to be tested (as this set cannot be minimal). Tests for invariance can be computationally expensive if one has large amounts of data. Therefore, skipping unnecessary tests offers a significant speedup. In the extreme case, where all singletons are found to be invariant, the algorithm completes in $d + 1$ steps, compared to $\sum_{i=0}^m \binom{d}{i}$ steps ($2^d$ if $m = d$). This is implemented in lines 8-10 of \cref{appendix:S_AS_algorithm_alg}.
\begin{algorithm}[h]
\caption{An algorithm for computing $\hat{S}_{\AS}$ from data}
\label{appendix:S_AS_algorithm_alg}
\begin{algorithmic}[1]
\INPUT{A decision rule $\phi_n$ for invariance, significance thresholds $\alpha_0, \alpha$, max size of sets to test $m$ (potentially $m = d$) and data}
\OUTPUT{The set $\hat{S}_{\AS}$}

\STATE Initialize $\widehat{\MIP}$ as an empty list.
\STATE $PS \gets \{S \subseteq [d] \mid \vert S \vert \leq m \}$ 
\IF{$\phi_n(\emptyset, \alpha_0) = 0$}
    \STATE End the procedure and return $\hat{S}_{\AS} = \emptyset$
\ENDIF
\STATE Sort $PS$ in increasing order according the set sizes
\FOR{$S \in PS$}
    \IF{$S \supsetneq S'$ for any $S' \in \widehat{\MIP}$}
        \STATE Skip the test of $S$ and go to next iteration of the loop
    \ELSE
        \STATE Add $S$ to $\widehat{\MIP}$ if $\phi_n(S, \alpha) = 0$, else continue
    \ENDIF
    \IF{The union of $\widehat{\MIP}$ contains all nodes}
        \STATE Break the loop
    \ENDIF
\ENDFOR
\STATE Return $\hat{S}_{\AS}$ as the union of all sets in $\widehat{\MIP}$
\end{algorithmic}
\end{algorithm}

\section{Additional Experiment Details}
\subsection{Simulation Details for \cref{experiment:1}}\label{appendix:graph_sampling}
We sample graphs that satisfy \cref{assumption:1} with the additional requirement that $Y \in \DE_Y$ by the following procedure:
\begin{enumerate}
    \item Sample a DAG $\mathcal{G}$ for the graph of $(X, Y)$ with $d + 1$ nodes, for $d \in \{4, 6, \dots, 20\} \cup \{100, 1{,}000\}$, and choose $Y$ to be a node (chosen uniformly at random) that is not a root node.
    \item Add a root node $E$ to $\mathcal{G}$ with $N_{\textrm{interventions}}$ children that are not $Y$. When $d \leq 20$, $N_{\textrm{interventions}} \in \{1, \dots, d\}$ and when $d \geq 100$, $N_{\textrm{interventions}} \in \{1, \dots, 0.1 \times d\}$ (i.e., we consider interventions on up to ten percent of the predictor nodes). 
    \item Repeat the first two steps if $Y \not\in \DE_E$.
\end{enumerate}

\subsection{Simulation Details for \cref{sec:experiment2}}\label{app:simulation_details}
We simulate data for the experiment in \cref{sec:experiment2} (and the additional plots in \cref{appendix:alpha0}) by the following procedure:
\begin{enumerate}
    \item Sample data from a single graph by the following procedure:
    \begin{enumerate}
        \item Sample a random graph $\mathcal G$ of size $d + 1$ and sample $Y$ (chosen uniformly at random) as any node that is not a root node in this graph. 
        \item Sample coefficients, $\beta_{i\rightarrow j}$, for all edges $(i \rightarrow j)$ in $\mathcal{G}$ from $U((-2, 0.5) \cup (0.5, 2))$ independently.
        \item Add a node $E$ with no incoming edges and $N_{\textrm{interventions}}$ children, none of which are $Y$. When $d = 6$, we set $N_{\textrm{interventions}} = 1$ and when $d = 100$, we sample $N_{\textrm{interventions}}$ uniformly from $\{1, \dots, 10\}$.
        \item If $Y$ is not a descendant of $E$, repeat steps (a), (b) and (c) until a graph where $Y \in \DE_E$ is obtained.
        \item For $n \in \{10^2, 10^3, 10^4, 10^5\}$:
            \begin{enumerate}
                \item Draw $50$ datasets of size $n$ from an SCM with graph $\mathcal{G}$ and coefficients $\beta_{i\rightarrow j}$ and with i.i.d.\ $N(0, 1)$ noise innovations. The environment variable, $E$, is sampled independently from a Bernoulli distribution with probability parameter $p = 0.5$, corresponding to (roughly) half the data being observational and half the data interventional. The data are generated by looping through a causal ordering of $(X, Y)$, starting at the bottom, and standardizing a node by its own empirical standard deviation before generating children of that node; that is, a node $X_j$ is first generated from $\PA_j$ and then standardized before generating any node in $\CH_j$. If $X_j$ is intervened on, we standardize it prior to the intervention.
                
                \item For each sampled dataset, apply IAS and ICP. Record the Jaccard similarities between IAS and $\AN_Y$ and between ICP and $\AN_Y$, and record whether or not is was a subset of $\AN_Y$ and whether it was empty. 
                
                \item Estimate the quantity plotted (average Jaccard similarity in \cref{fig:experiment2_jaccard} or probability of $\hat{S}_{\AS} \subseteq \AN_Y$ or $\hat{S}_{\AS} = \emptyset$ in \cref{fig:appendix_exp2_main}) from the $50$ simulated datasets.
            \end{enumerate}
        \item Return the estimated quantities from the previous step.
    \end{enumerate}
    \item Repeat the above $100$ times and save the results in a data-frame.
\end{enumerate}

\subsection{Analysis of the Choice of $C$ in \cref{sec:experiment2}}\label{appendix:correction_level}
We have repeated the simulation with $d = 6$ from \cref{sec:experiment2} but with a correction factor of $C = 2^6$, as suggested by \cref{theorem:exhaustive_search} instead of the heuristic correction factor of $C = 9$ suggested in \cref{appendix:simulate_max}. 
\cref{fig:appendix_exp2_correctionfactor} shows the results.
We see that the results are almost identical to those presented in \cref{fig:experiment2_jaccard}. Thus, in the scenario considered here, there is no change in the performance of $\hat{S}_{\AS}$ (as measured by Jaccard similarity) between using a correction factor of $C = 2^6$ and a correction factor of $C = 3^{\lceil 6/3\rceil} = 9$. In larger graphs, it is likely that there is a more pronounced difference. E.g., at $d = 10$, the strictly conservative correction factor suggested by \cref{theorem:exhaustive_search} is $2^{10} = 1024$, whereas the correction factor suggested in \cref{appendix:simulate_max} is only $3^{\lceil 10/3 \rceil} = 3^4 = 81$, and at $d = 20$ the two are $2^{20} = 1{,}048{,}576$ and $3^{\lceil 20/3 \rceil} = 3^7 = 2187$.
\begin{figure}[ht]
    \centerline{
    \includegraphics[width=.6\linewidth]{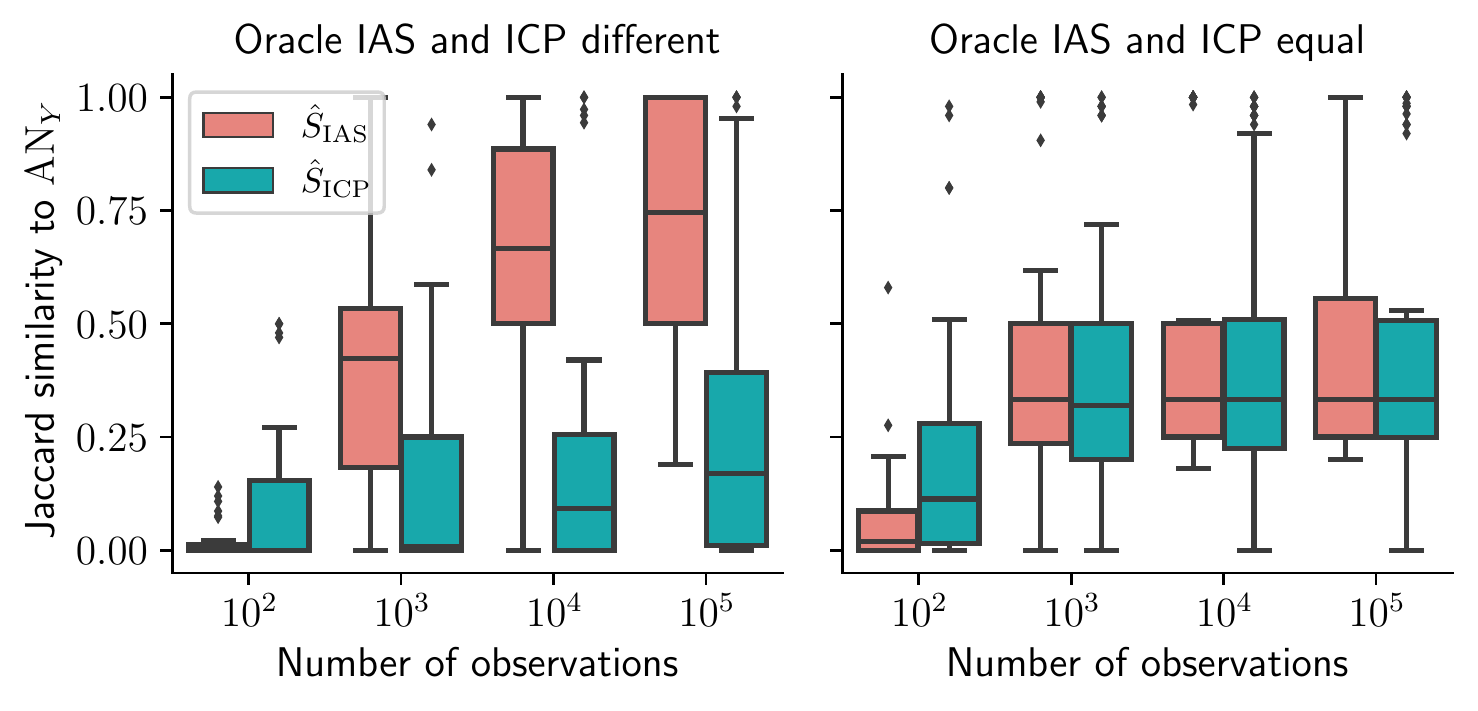}
    }
    \caption{The same figure as in \cref{fig:experiment2_jaccard}, but with a correction factor of $C = 2^6 = 64$ instead of $C = 3^{\lceil 6 / 3\rceil} = 9$. Only $d = 6$ shown here, as the correction factor for $d = 100$ is unchanged. Here, the guarantees of \cref{theorem:exhaustive_search} are not violated by a potentially too small correction factor, and the results are near identical to those given in \cref{fig:experiment2_jaccard} using a milder correction factor.
    }
    \label{fig:appendix_exp2_correctionfactor}
\end{figure}

\subsection{Analysis of the Choice of $\alpha_0$ in \cref{sec:experiment2}}\label{appendix:alpha0}
\begin{figure}[ht]
    \centerline{
    \includegraphics[width=8cm]{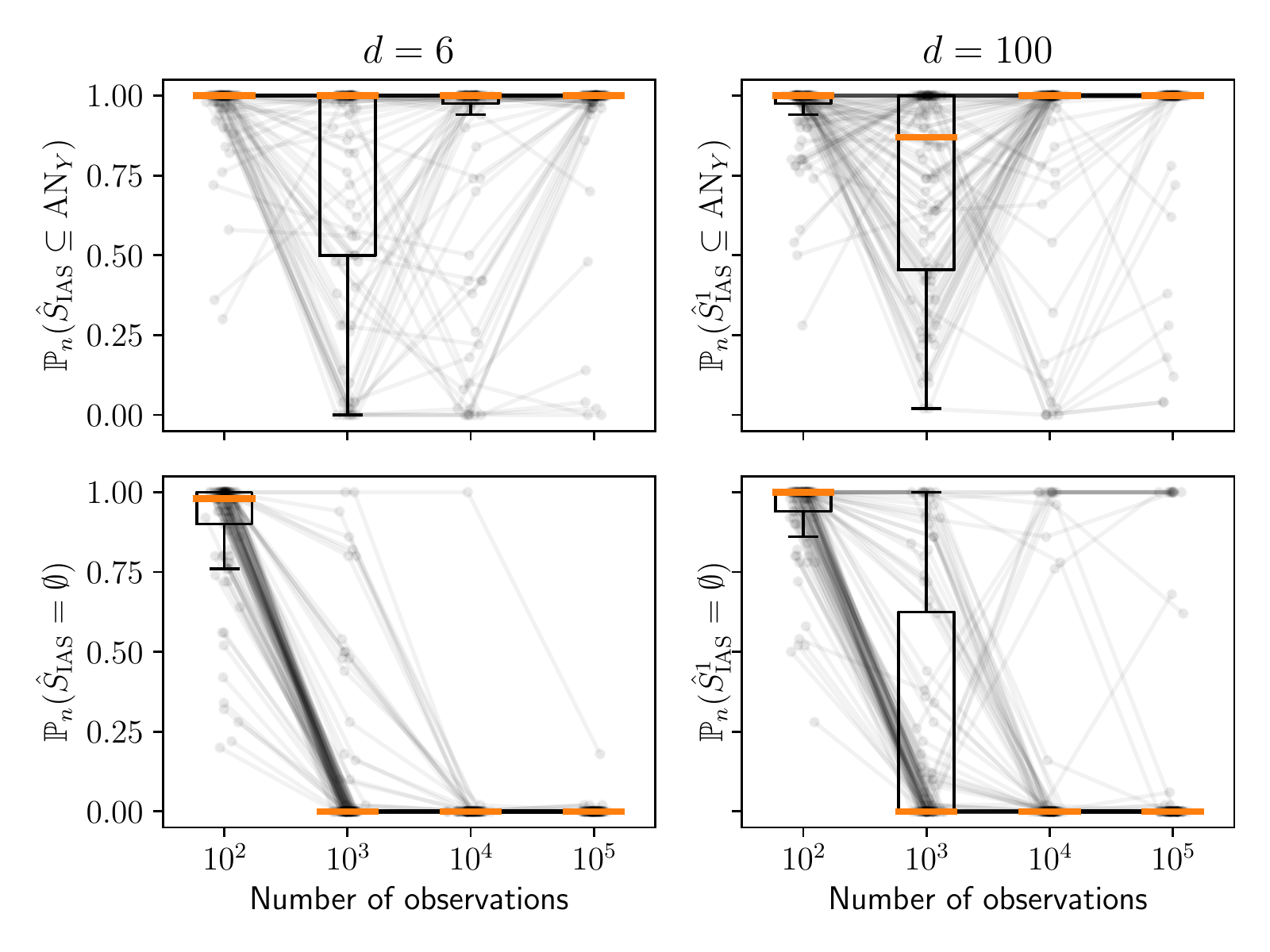}
    }
    \caption{
    The empirical probabilities of recovering a subset of $\AN_Y$ (top row) and recovering an empty set (bottom row), when testing the empty set for invariance at level $\alpha_0 = 10^{-6}$. Generally, our methods seem to hold level well, especially when sample sizes are large. When the sample size is small, the output is often the empty set.
    When $d = 6$, we estimate $\hat{S}_{\AS}$ (left column) and when $d = 100$, we estimate $\hat{S}_{\AS}^1$ (right column). The results here are from the simulations that also produced \cref{fig:experiment2_jaccard}.
    Medians are displayed as orange lines through each boxplot.
    Each point represents the probability that the output set is ancestral (resp. empty) for a randomly selected SCM, as estimated by repeatedly sampling data from the same SCM for every $n \in \{10^2, 10^3, 10^4, 10^5\}$. Observations from the same SCM are connected by a line. Each figure contains data from $100$ randomly drawn SCMs. Points have been perturbed slightly along the $x$-axis to improve readability. 
    }
    \label{fig:appendix_exp2_main}
\end{figure}

\begin{figure}[ht]
    \centerline{
    \includegraphics[width=8cm]{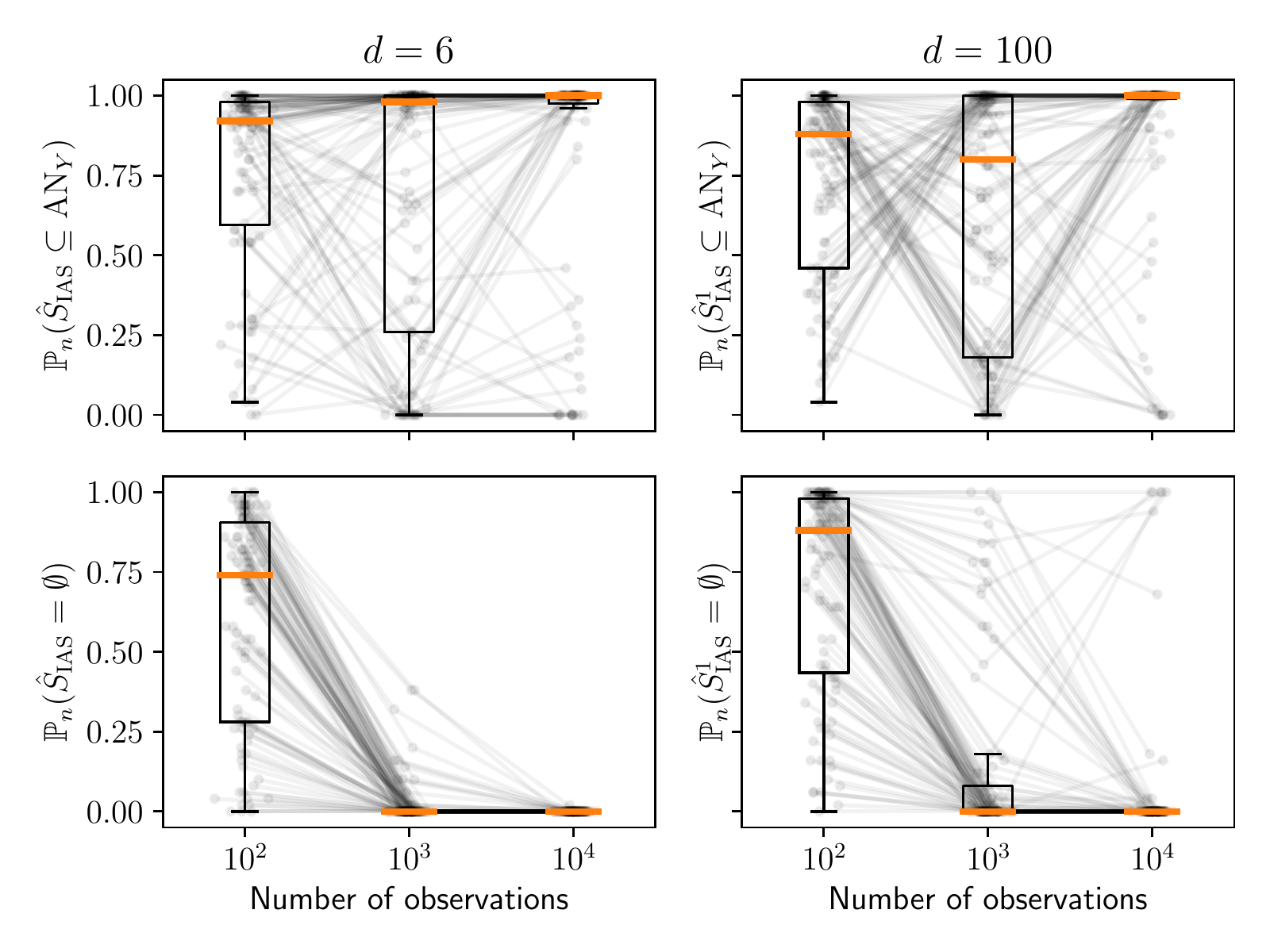}
    }
    \caption{The same figure as \cref{fig:appendix_exp2_main}, but with $\alpha_0 = \alpha = 0.05$ and $n \in \{10^2, 10^3, 10^4\}$. Testing the empty set at the non-conservative level $\alpha_0 = \alpha$ means that the empty set is output less often for small sample sizes, but decreases the probability that the output is a subset of ancestors. Thus, we find more ancestor-candidates, but make more mistakes when $\alpha_0 = \alpha$. However, the median probability of the output being a subset of ancestors is at least $80\%$ in all configurations.
    }
    \label{fig:appendix_exp2_None}
\end{figure}

\begin{figure}[ht]
    \centerline{
    \includegraphics[width=8cm]{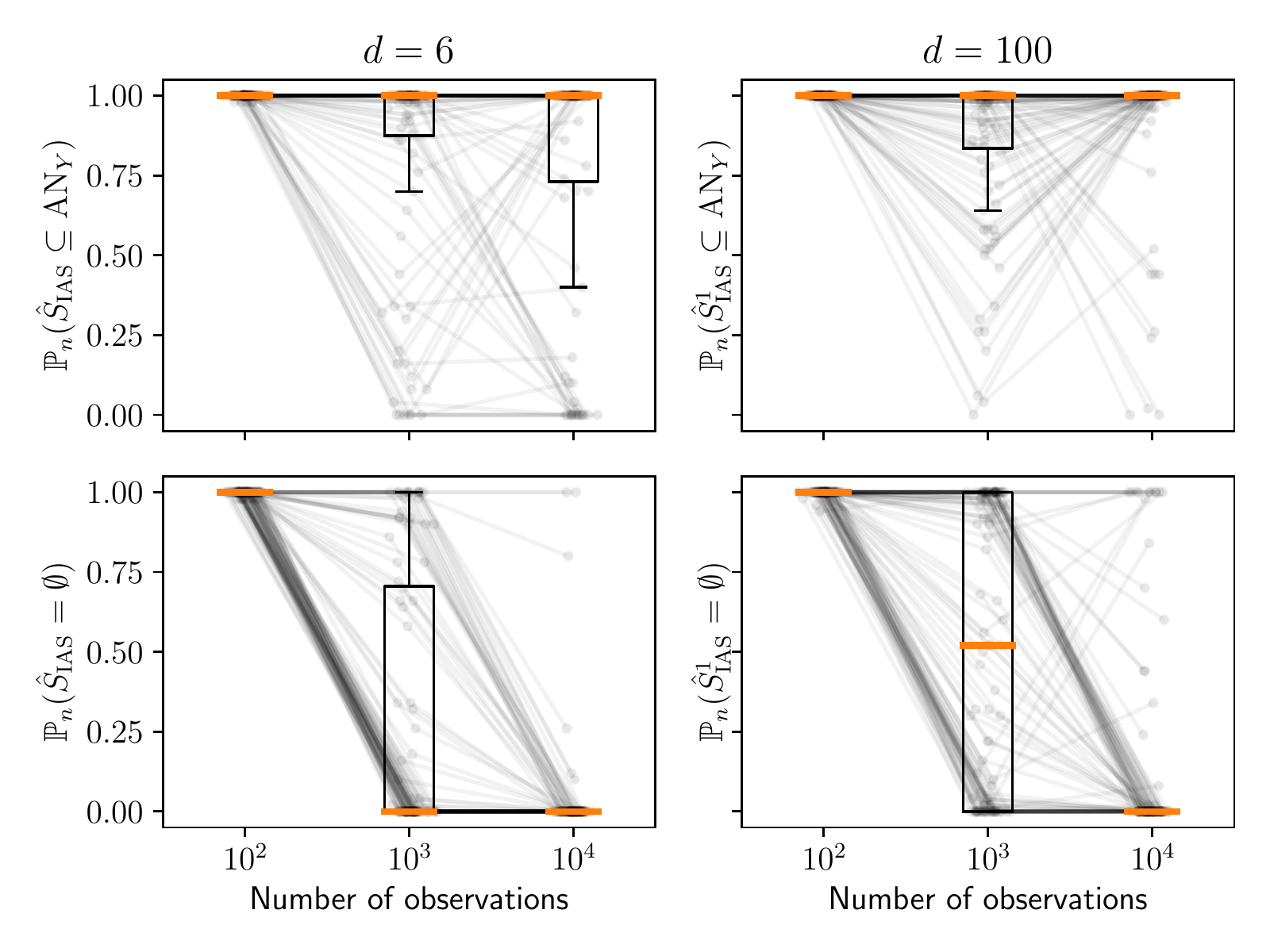}
    }
    \caption{The same figure as \cref{fig:appendix_exp2_main}, but with $\alpha_0 = 10^{-12}$ and $n \in \{10^2, 10^3, 10^4\}$. Testing the empty set at at very conservative level $\alpha_0 = 10^{-12}$ means that the empty set is output more often (for one hundred observations, we only find the empty set), but increases the probability that the output is a subset of ancestors. Thus, testing at a very conservative level $\alpha_0 = 10^{-12}$ means that we do not make many mistakes, but the output is often non-informative. 
    }
    \label{fig:appendix_exp2_minu12}
\end{figure}

Here, we investigate the quantities $\mathbb{P}(\hat{S}_{\AS} \subseteq \AN_Y)$, $\mathbb{P}(\hat{S}_{\AS}^1 \subseteq \AN_Y)$, $\mathbb{P}(\hat{S}_{\AS} = \emptyset)$ and $\mathbb{P}(\hat{S}_{\AS}^1 = \emptyset)$  
using the same simulation setup as described in \cref{sec:experiment2}. Furthermore, we also ran the simulations for values $\alpha_0 = \alpha$ (testing all hypotheses at the same level), $\alpha_0 = 10^{-6}$ (conservative, see \cref{remark:alpha0}) as in \cref{sec:experiment2}  and $\alpha_0 = 10^{-12}$ (very conservative). 
The results for $\alpha = 10^{-6}$ (shown in \cref{fig:appendix_exp2_main}) were recorded in the same simulations that produced the output for \cref{fig:experiment2_jaccard}. For $\alpha_0 \in \{\alpha, 10^{-12}\}$ (shown in \cref{fig:appendix_exp2_None} and \cref{fig:appendix_exp2_minu12}, respectively) we only simulated up to $10{,}000$ observations, to keep computation time low.

Generally, we find that the probability of IAS being a subset of the ancestors 
seems to generally hold well and even more so with large sample sizes.
(see \cref{fig:appendix_exp2_main,fig:appendix_exp2_None,fig:appendix_exp2_minu12}), in line with \cref{theorem:exhaustive_search}. When given $100{,}000$ observations, the probability of IAS being a subset of ancestors is roughly equal to one for almost all SCMs, although there are a few SCMs, where IAS is never a subset of the ancestors (see \cref{fig:appendix_exp2_main}). For $\alpha_0 = 10^{-6}$, the median probability of IAS containing only ancestors is one in all cases, except for $d = 100$ with $1{,}000$ observations -- here, the median probability is $87\%$.

In general, varying $\alpha_0$ has the effect hypothesized in \cref{remark:alpha0}: lowering $\alpha_0$ increases the probability that IAS contains only ancestors, but at the cost of increasing the probability that it is empty (see \cref{fig:appendix_exp2_main,fig:appendix_exp2_None,fig:appendix_exp2_minu12}). For instance, the median probability of IAS being a subset of ancestors when $\alpha_0 = 10^{-12}$ is one for all sample sizes, but the output is always empty when there are $100$ observations and empty roughly half the time even at $1{,}000$ observations when $d = 100$ (see \cref{fig:appendix_exp2_minu12}). In contrast, not testing the empty set at a reduced level, means that the output of IAS is rarely empty, but the probability of IAS containing only ancestors decreases. Still, even with $\alpha_0 = \alpha$, the median probability of IAS containing only ancestors was never lower than $80\%$ (see \cref{fig:appendix_exp2_None}). Thus, choosing $\alpha_0$ means choosing a trade-off between finding more ancestor-candidates, versus more of them being false positives.

\subsection{Analysis of the strength of inverventions in \cref{sec:experiment2}}\label{appendix:weakInstrument}
Here, we repeat the $d = 6$ simulations from \cref{sec:experiment2} with a reduced strength of the environment to investigate the performance of IAS under weaker interventions. We sample from the same SCMs as sampled in \cref{sec:experiment2}, but reduce the strength of the interventions to be $0.5$ instead of $1$. That is, the observational distributions are the same as in \cref{sec:experiment2}, but interventions to a node $X_j$ are here half as strong as in \cref{sec:experiment2}. 

\begin{figure}[ht]
    \centerline{
    \includegraphics[width=.6\linewidth]{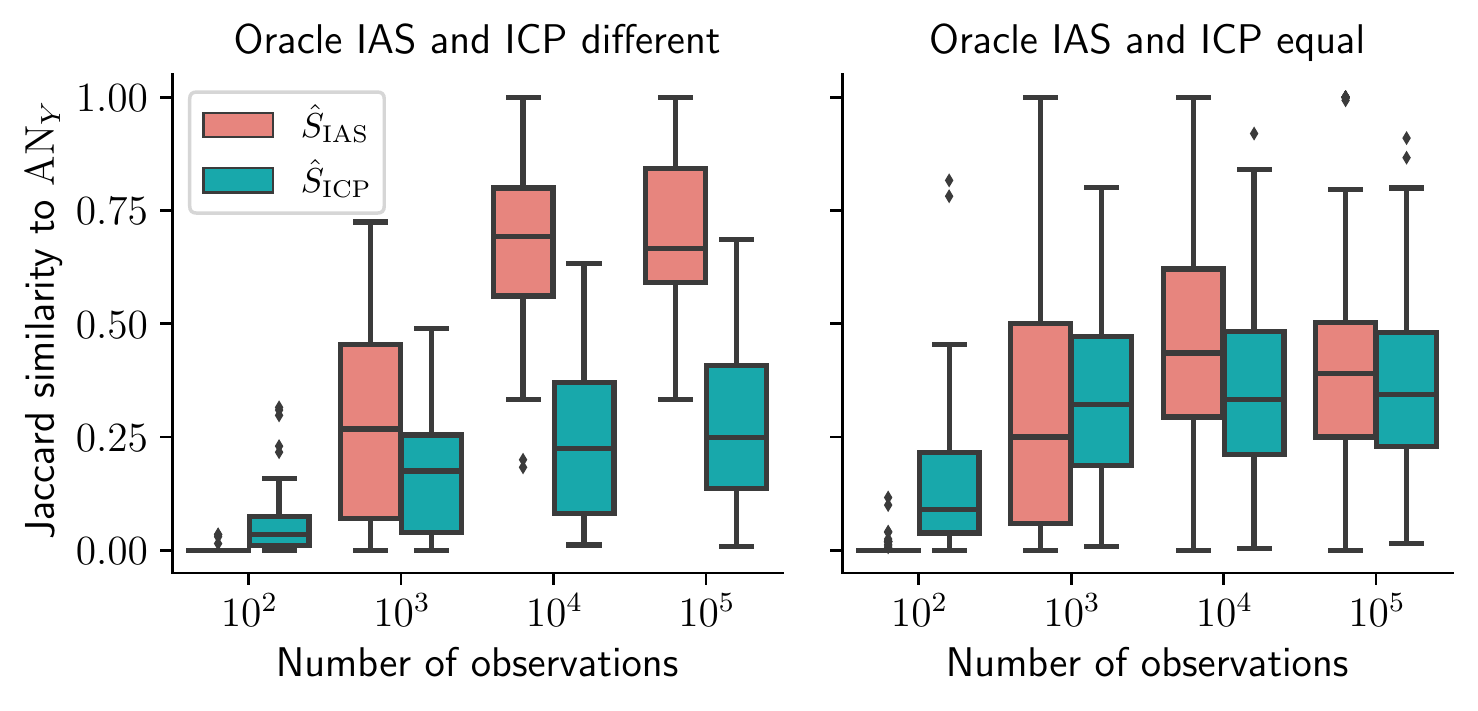}
    }
    \caption{
    The same figure as the one presented in \cref{fig:experiment2_jaccard}, but with weaker environments (do-interventions of strength $0.5$ compared to $1$ in \cref{fig:experiment2_jaccard}). Generally, IAS performs the same for weaker interventions as for strong interventions, when there are more than $10{,}000$ observations.
    Graphs represented in each boxplot: $42$ (left), $58$ (right).
    }
    \label{fig:appendix_jaccard_weak}
\end{figure}

The Jaccard similarity between $\hat{S}_{\AS}$ and $\AN_Y$ is generally lower than what we found in \cref{fig:experiment2_jaccard} (see \cref{fig:appendix_jaccard_weak}). This is likely due to having lower power to detect non-invariance, which has two implications. First, lower power means that we may fail to reject the empty set, meaning that we output nothing. Then, the Jaccard similarity between $\hat{S}_{\AS}$ and $\AN_Y$ is zero. Second, it may be that we correctly reject the empty set, but fail to reject another non-invariant set which is not an ancestor of $Y$ which is then potentially included in the output. Then, the $\hat{S}_{\AS}$ and $\AN_Y$ is lower, because we increase the number of false findings.

We find that the probability that $\hat{S}_{\AS}$ is a subset of ancestors is generally unchanged for the lower intervention strength, but the probability of $\hat{S}_{\AS}$ generally increases for small sample sizes (see \cref{appendix:table_probs}). This indicates that IAS does not make more mistakes under the weaker interventions, but it is more often uninformative. We see also that in both settings, $\hat{S}_{\AS}$ is empty more often than $\hat{S}_{\ICP}$ for low sample sizes, but less often for larger samples (see \cref{appendix:table_probs}). This is likely because IAS tests the empty set at a much lower level than ICP does ($10^{-6}$ compared to $0.05$). Thus, IAS requires more power to find anything, but once it has sufficient power, it finds more than ICP (see also \cref{fig:appendix_jaccard_weak}). The median probability of ICP returning a subset of the ancestors was always at least $95\%$ (not shown).

$\text{ } $ \hfill \\
$\text{ } $
\newpage

\begin{table}[ht]
\caption{Summary of the quantities $\mathbb{P}(\hat{S}_{\AS} \subseteq \AN_Y)$, $\mathbb{P}(\hat{S}_{\AS} = \emptyset)$ and $\mathbb{P}(\hat{S}_{\ICP} = \emptyset)$ for weak and strong do-interventions (strength $0.5$ and $1$, respectively) when $d = 6$. Numbers not in parentheses are means, numbers in parentheses are medians. The level is generally unchanged when the environments have a weaker effect, but the power is lower, in the sense that the empty set is output more often.}
\label{appendix:table_probs}
\centerline{
\begin{tabular}{lllll}
                                      &                                      & $\mathbb{P}(\hat{S}_{\AS} \subseteq \AN_Y)$ & $\mathbb{P}(\hat{S}_{\AS} = \emptyset)$ & $\mathbb{P}(\hat{S}_{\ICP} = \emptyset)$ \\ \hline
\multirow{4}{*}{Strong interventions} & \multicolumn{1}{l|}{$n = 100$}       & 96.6\% (100\%)                              & 89.6\% (98\%)                           & 52.3\% (52\%)                            \\
                                      & \multicolumn{1}{l|}{$n = 1{,}000$}   & 75.7\% (100\%)                              & 10.0\% (0\%)                            & 30.4\% (14\%)                            \\
                                      & \multicolumn{1}{l|}{$n = 10{,}000$}  & 83.7\% (100\%)                              & 1.0\% (0\%)                             & 24.9\% (10\%)                            \\
                                      & \multicolumn{1}{l|}{$n = 100{,}000$} & 93.8\% (100\%)                              & 0.2\% (0\%)                                  & 22.9\% (10\%)                            \\ \hline
\multirow{4}{*}{Weak interventions}   & \multicolumn{1}{l|}{$n = 100$}       & 99.3\% (100\%)                              & 98.7\% (100\%)                          & 72.0\% (84\%)                            \\
                                      & \multicolumn{1}{l|}{$n = 1{,}000$}   & 81.1\% (100\%)                              & 40.2\% (26\%)                           & 36.9\% (24\%)                            \\
                                      & \multicolumn{1}{l|}{$n = 10{,}000$}  & 80.8\% (100\%)                              & 1.7\% (0\%)                             & 27.5\% (15\%)                            \\
                                      & \multicolumn{1}{l|}{$n = 100{,}000$} & 92.6\% (100\%)                              & 1.1\% (0\%)                             & 24.8\% (14\%)                           
\end{tabular}}
\end{table}

$\text{ } $ \hfill \\
$\text{ } $
\newpage

\subsection{Analysis of the Choice of $q_{TB}$ in \cref{sec:experiment_gene}}\label{appendix:q-tb}
\begin{figure}[ht]
    \centering
    \includegraphics[width=0.5\linewidth]{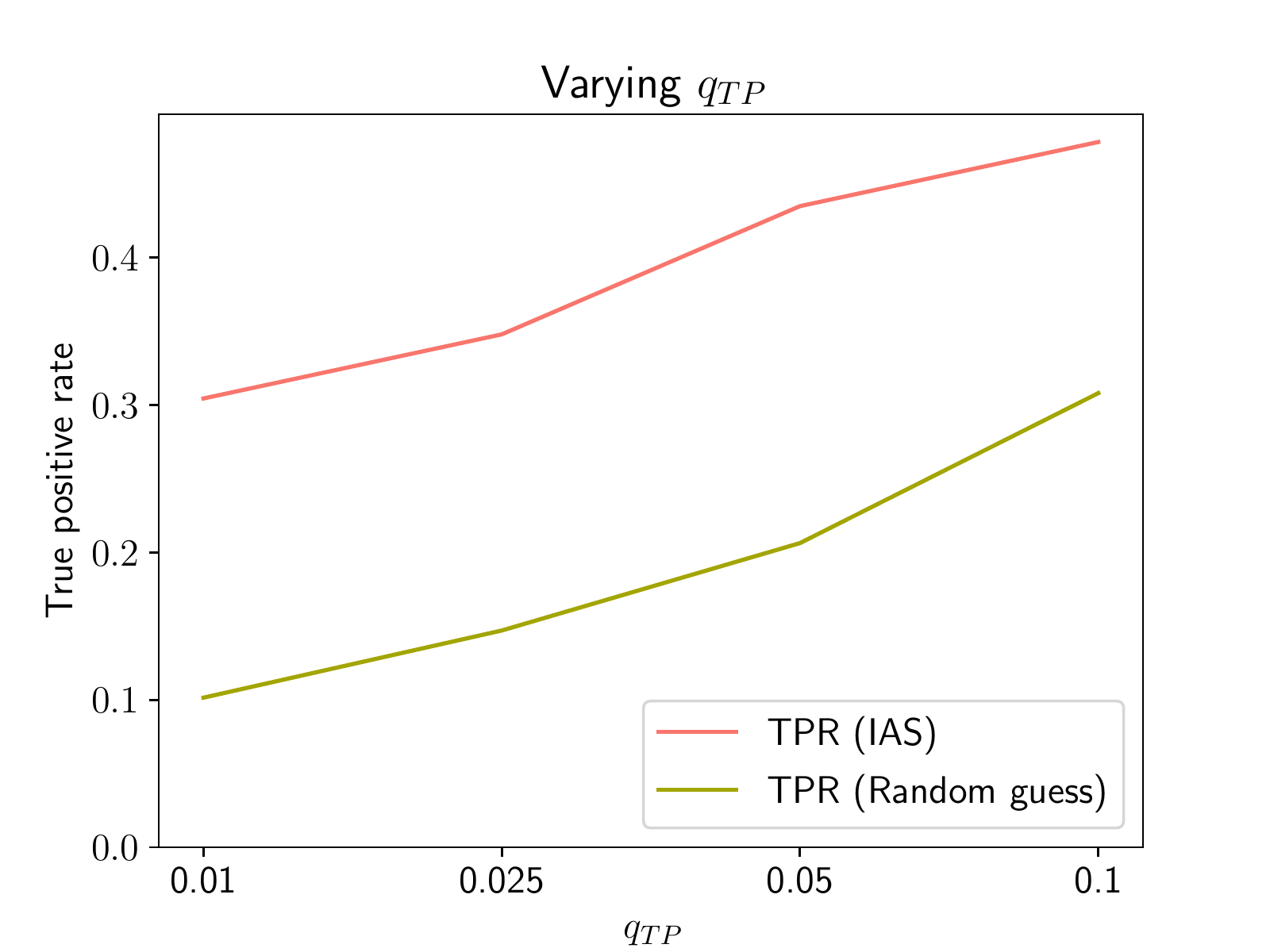}
    \caption{True positive rates (TPRs) for the gene experiment in \cref{sec:experiment_gene}. $q_{TB}$ specifies the quantile in the observed distribution that an intervention effect has to exceed to be considered a true positive. 
    While the TPR increases for our method when $q_{TB}$ is increased, the TPR of random guessing increases comparably. This validates that 
    changing the definition of true positives in this experiment by choosing a different  $q_{TB}$ does not change the conclusion of the experiment substantially.
    }
    \label{fig:changing-q-tb}
\end{figure}

In this section, we analyze the effect of changing the cut-off $q_{TB}$ that determines when a gene pair is considered a true positive in \cref{sec:experiment_gene}. For the results in the main paper, we use $q_{TB} = 1\%$, meaning that the pair $(\gene_X, \gene_Y)$ is considered a true positive if the value of $\gene_Y$ when intervening on $\gene_X$ is outside of the $0.01$- and $0.99$-quantiles of $\gene_Y$ in the observational distribution. 
In \cref{fig:changing-q-tb}, we plot the true positive rates for several other choices of $q_{TB}$. We compare to the true positive rate of random guessing, which also increases if the criterion becomes easier to satisfy.
We observe that the choice of $q_{TB}$ does not substantially change the excess true positive rate of our method compared to random guessing. This indicates that while the true positives in this experiments are inferred from data, the conclusions drawn in \cref{fig:experiment-gene} are robust with respect to some modelling choices of $q_{TB}$.

\subsection{Learning causal ancestors by estimating the I-MEC}\label{app:UT-IGSP}
\begin{figure}[ht]
    \centerline{
    \includegraphics[width=.6\linewidth]{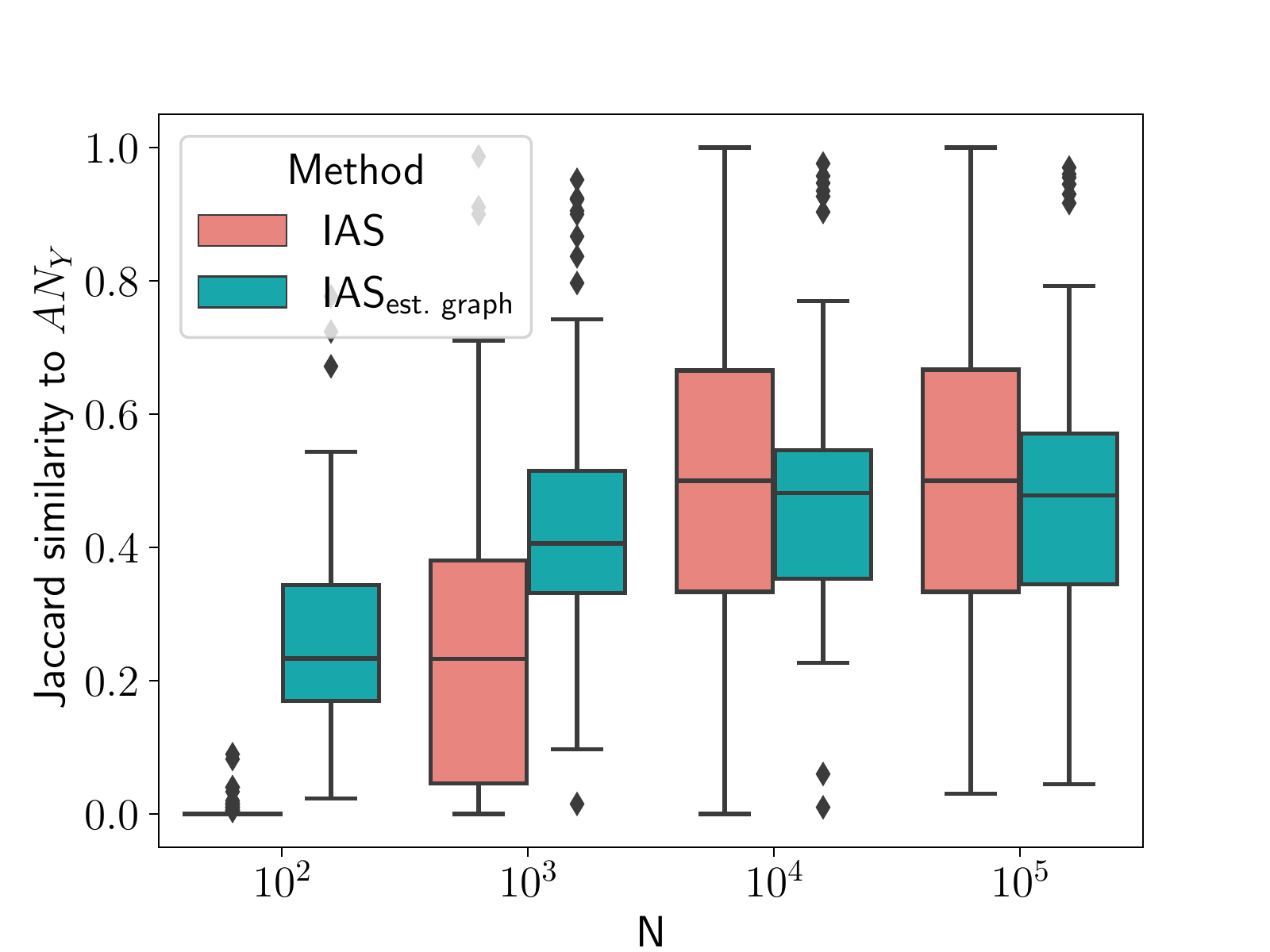}
    }
    \caption{
    Comparison between the finite sample output of IAS and the procedure described in \cref{app:UT-IGSP}, in the low-dimensional case. Generally, these procedures have similar performance, although IAS performs worse for small sample sizes but slightly better for high sample sizes.
    }
    \label{fig:appendix_UT-IGSP}
\end{figure}

In this section, we repeat the experiments performed in \cref{sec:experiment2}, this time including a procedure (here denoted \IASest), where we perform the following steps.
    \begin{enumerate}
        \item Estimate a member graph of the I-MEC and the location of the intervention sites using Unknown-Target Interventional Greedy Sparsest Permutation (UT-IGSP) \citep{squires2020permutation} using the implemention from the Python package CausalDAG.\footnote{Available at \url{https://github.com/uhlerlab/causaldag}.}
        
        \item Apply the oracle algorithm described in \cref{sec:oracle-algorithm} to the estimated graph to obtain an estimate of $\MIP$.
        
        \item Output the union of all sets in the estimate of $\MIP$.
    \end{enumerate}
The results for the low-dimensional experiment are displayed in \cref{fig:appendix_UT-IGSP} and the results for the high-dimensional experiment are displayed in \cref{tab:appendix_UT-IGSP}. Here, we see that {\IASest} generally performs well (as measured by Jaccard similarity) in the low-dimensional setting ($d = 6$), and even better than IAS for sample sizes $N \leq 10^3$, but is slightly outperformed by IAS for larger sample sizes. 
However, in the high-dimensional setting ($d = 100$), we observe that {\IASest} fails to hold level and identifies only very few ancestors (see \cref{tab:appendix_UT-IGSP}). We hypothesize that the poor performance of {\IASest} in the high-dimensional setting is due to {\IASest} attempting to solve a more difficult task than IAS. {\IASest} first estimates
a full graph (here using UT-IGSP), even though only a subgraph of the full graph is of relevance in this scenario. In addition, UT-IGSP 
aims to
estimate the site of the unknown interventions. In contrast, 
IAS only needs to identify nodes that are capable of blocking all paths between two variables, and does not need to know the site of the interventions.

\begin{table}[ht]
    \centering
\begin{tabular}{lllllll}
                                                             & \multicolumn{2}{c}{$d = 100, N = 10^3$}    & \multicolumn{2}{c}{$d = 100,N = 10^4$}     & \multicolumn{2}{c}{$d = 100,N = 10^5$} \\
                                                             & IAS       & \IASest                        & IAS       & \IASest                        & IAS                & \IASest           \\ \cline{2-7} 
\multicolumn{1}{l|}{$\mathbb{P}(S_{\cdot} \subseteq \AN_Y)$} & $84.64\%$ & \multicolumn{1}{l|}{$15.30\%$} & $94.04\%$ & \multicolumn{1}{l|}{$14.92\%$} & $94.72\%$          & $14.74\%$         \\
\multicolumn{1}{l|}{$\mathbb{P}(S_{\cdot} = \emptyset)$}     & $51.96\%$ & \multicolumn{1}{l|}{$12.32\%$} & $12.72\%$ & \multicolumn{1}{l|}{$11.84\%$} & $6.98\%$           & $11.42\%$         \\
\multicolumn{1}{l|}{$J(S_{\cdot}, \AN_Y)$}                   & $0.19$    & \multicolumn{1}{l|}{$0.10$}    & $0.33$    & \multicolumn{1}{l|}{$0.10$}    & $0.35$             & $0.11$           
\end{tabular}
    \caption{Identifying ancestors by first estimating the I-MEC of the underlying DAG and then applying the oracle algorithm of \cref{sec:oracle-algorithm} fails to hold level and identifies fewer ancestors than applying IAS, when in a high-dimensional setting.}
    \label{tab:appendix_UT-IGSP}
\end{table}


\clearpage
\begin{thebibliography}{41}
\providecommand{\natexlab}[1]{#1}
\providecommand{\url}[1]{\texttt{#1}}
\expandafter\ifx\csname urlstyle\endcsname\relax
  \providecommand{\doi}[1]{doi: #1}\else
  \providecommand{\doi}{doi: \begingroup \urlstyle{rm}\Url}\fi

\bibitem[Acid \& De~Campos(2013)Acid and De~Campos]{acid2013algorithm}
Acid, S. and De~Campos, L.~M.
\newblock An algorithm for finding minimum d-separating sets in belief
  networks.
\newblock In \emph{Proceedings of the 29th Annual Conference on {U}ncertainty
  in {A}rtificial {I}ntelligence ({UAI})}, 2013.

\bibitem[Arjovsky et~al.(2019)Arjovsky, Bottou, Gulrajani, and
  Lopez-Paz]{arjovsky2019invariant}
Arjovsky, M., Bottou, L., Gulrajani, I., and Lopez-Paz, D.
\newblock Invariant risk minimization.
\newblock \emph{arXiv preprint arXiv:1907.02893}, 2019.

\bibitem[Berrett et~al.(2020)Berrett, Wang, Barber, and
  Samworth]{berrett2020conditional}
Berrett, T.~B., Wang, Y., Barber, R.~F., and Samworth, R.~J.
\newblock The conditional permutation test for independence while controlling
  for confounders.
\newblock \emph{Journal of the Royal Statistical Society: Series B (Statistical
  Methodology)}, 82\penalty0 (1):\penalty0 175--197, 2020.

\bibitem[Bongers et~al.(2021)Bongers, Forr{\'e}, Peters, and
  Mooij]{bongers2021foundations}
Bongers, S., Forr{\'e}, P., Peters, J., and Mooij, J.~M.
\newblock Foundations of structural causal models with cycles and latent
  variables.
\newblock \emph{Annals of Statistics}, 49\penalty0 (5):\penalty0 2885--2915,
  2021.

\bibitem[Chickering(2002)]{chickering2002optimal}
Chickering, D.~M.
\newblock Optimal structure identification with greedy search.
\newblock \emph{Journal of Machine Learning Research}, 3:\penalty0 507--554,
  2002.

\bibitem[Christiansen et~al.(2022)Christiansen, Pfister, Jakobsen, Gnecco, and
  Peters]{christiansen2021causal}
Christiansen, R., Pfister, N., Jakobsen, M.~E., Gnecco, N., and Peters, J.
\newblock A causal framework for distribution generalization.
\newblock \emph{IEEE Transactions on Pattern Analysis and Machine Intelligence
  (accepted)}, 2022.

\bibitem[Friedman(2001)]{friedman2001greedy}
Friedman, J.~H.
\newblock Greedy function approximation: a gradient boosting machine.
\newblock \emph{Annals of Statistics}, 29\penalty0 (5):\penalty0 1189--1232,
  2001.

\bibitem[Fukumizu et~al.(2008)Fukumizu, Gretton, Sun, and
  Sch\"olkopf]{Fukumizu2008}
Fukumizu, K., Gretton, A., Sun, X., and Sch\"olkopf, B.
\newblock Kernel measures of conditional dependence.
\newblock In \emph{{A}dvances in {N}eural {I}nformation {P}rocessing {S}ystems
  ({NeurIPS})}, volume~20, 2008.

\bibitem[Gamella \& Heinze-Deml(2020)Gamella and
  Heinze-Deml]{gamella2020active}
Gamella, J.~L. and Heinze-Deml, C.
\newblock Active invariant causal prediction: Experiment selection through
  stability.
\newblock In \emph{{A}dvances in {N}eural {I}nformation {P}rocessing {S}ystems
  ({NeurIPS})}, volume~33, 2020.

\bibitem[Gaspers \& Mackenzie(2015)Gaspers and Mackenzie]{gaspers2015number}
Gaspers, S. and Mackenzie, S.
\newblock On the number of minimal separators in graphs.
\newblock In \emph{International Workshop on Graph-Theoretic Concepts in
  Computer Science}, pp.\  116--121. Springer, 2015.

\bibitem[Heinze-Deml et~al.(2018)Heinze-Deml, Peters, and
  Meinshausen]{heinze2018invariant}
Heinze-Deml, C., Peters, J., and Meinshausen, N.
\newblock Invariant causal prediction for nonlinear models.
\newblock \emph{Journal of Causal Inference}, 6\penalty0 (2), 2018.

\bibitem[Hoyer et~al.(2009)Hoyer, Janzing, Mooij, Peters, and
  Sch\"{o}lkopf]{hoyer2008nonlinear}
Hoyer, P., Janzing, D., Mooij, J.~M., Peters, J., and Sch\"{o}lkopf, B.
\newblock Nonlinear causal discovery with additive noise models.
\newblock In \emph{{A}dvances in {N}eural {I}nformation {P}rocessing {S}ystems
  ({NeurIPS})}, volume~21, 2009.

\bibitem[Jordan(2019)]{jordan2019artificial}
Jordan, M.~I.
\newblock Artificial intelligence — the revolution hasn’t happened yet.
\newblock \emph{Harvard Data Science Review}, 1\penalty0 (1), 2019.

\bibitem[Kemmeren et~al.(2014)Kemmeren, Sameith, Van De~Pasch, Benschop,
  Lenstra, Margaritis, O’Duibhir, Apweiler, van Wageningen, Ko,
  et~al.]{kemmeren2014large}
Kemmeren, P., Sameith, K., Van De~Pasch, L.~A., Benschop, J.~J., Lenstra,
  T.~L., Margaritis, T., O’Duibhir, E., Apweiler, E., van Wageningen, S., Ko,
  C.~W., et~al.
\newblock Large-scale genetic perturbations reveal regulatory networks and an
  abundance of gene-specific repressors.
\newblock \emph{Cell}, 157\penalty0 (3):\penalty0 740--752, 2014.

\bibitem[Lauritzen(1996)]{lauritzen1996graphical}
Lauritzen, S.~L.
\newblock \emph{Graphical models}.
\newblock Clarendon Press, 1996.

\bibitem[Magliacane et~al.(2018)Magliacane, van Ommen, Claassen, Bongers,
  Versteeg, and Mooij]{Magliacane2018}
Magliacane, S., van Ommen, T., Claassen, T., Bongers, S., Versteeg, P., and
  Mooij, J.~M.
\newblock Domain adaptation by using causal inference to predict invariant
  conditional distributions.
\newblock In Bengio, S., Wallach, H., Larochelle, H., Grauman, K.,
  Cesa-Bianchi, N., and Garnett, R. (eds.), \emph{Advances in Neural
  Information Processing Systems 31}, pp.\  10846--10856. Curran Associates,
  Inc., 2018.

\bibitem[Martinet et~al.(2021)Martinet, Strzalkowski, and
  Engelhardt]{martinet2021variance}
Martinet, G., Strzalkowski, A., and Engelhardt, B.~E.
\newblock Variance minimization in the {W}asserstein space for invariant causal
  prediction.
\newblock \emph{arXiv preprint arXiv:2110.07064}, 2021.

\bibitem[Meinshausen \& B{\"u}hlmann(2006)Meinshausen and
  B{\"u}hlmann]{meinshausen2006high}
Meinshausen, N. and B{\"u}hlmann, P.
\newblock High-dimensional graphs and variable selection with the lasso.
\newblock \emph{Annals of Statistics}, 34\penalty0 (3):\penalty0 1436--1462,
  2006.

\bibitem[Mooij et~al.(2020)Mooij, Magliacane, and Claassen]{JCI}
Mooij, J.~M., Magliacane, S., and Claassen, T.
\newblock Joint causal inference from multiple contexts.
\newblock \emph{Journal of Machine Learning Research}, 21\penalty0
  (99):\penalty0 1--108, 2020.
\newblock URL \url{http://jmlr.org/papers/v21/17-123.html}.

\bibitem[Pearl(2009)]{pearl2009causality}
Pearl, J.
\newblock \emph{Causality}.
\newblock Cambridge university press, 2009.

\bibitem[Pearl(2014)]{PearlJudea2014Prii}
Pearl, J.
\newblock \emph{Probabilistic reasoning in intelligent systems: networks of
  plausible inference}.
\newblock The Morgan Kaufmann series in representation and learning. Morgan
  Kaufmann, 2014.

\bibitem[Pearl(2018)]{pearl2018theoretical}
Pearl, J.
\newblock Theoretical impediments to machine learning with seven sparks from
  the causal revolution.
\newblock \emph{arXiv preprint arXiv:1801.04016}, 2018.

\bibitem[Peters et~al.(2014)Peters, Mooij, Janzing, and
  Sch\"olkopf]{Peters2014jmlr}
Peters, J., Mooij, J.~M., Janzing, D., and Sch\"olkopf, B.
\newblock Causal discovery with continuous additive noise models.
\newblock \emph{Journal of Machine Learning Research}, 15:\penalty0 2009--2053,
  2014.

\bibitem[Peters et~al.(2016)Peters, B{\"u}hlmann, and
  Meinshausen]{peters2016causal}
Peters, J., B{\"u}hlmann, P., and Meinshausen, N.
\newblock Causal inference by using invariant prediction: identification and
  confidence intervals.
\newblock \emph{Journal of the Royal Statistical Society. Series B (Statistical
  Methodology)}, pp.\  947--1012, 2016.

\bibitem[Pfister et~al.(2019)Pfister, B{\"u}hlmann, and
  Peters]{pfister2019invariant}
Pfister, N., B{\"u}hlmann, P., and Peters, J.
\newblock Invariant causal prediction for sequential data.
\newblock \emph{Journal of the American Statistical Association}, 114\penalty0
  (527):\penalty0 1264--1276, 2019.

\bibitem[{R Core Team}(2021)]{Rsoftware}
{R Core Team}.
\newblock \emph{{R}: A Language and Environment for Statistical Computing}.
\newblock {R} Foundation for Statistical Computing, Vienna, Austria, 2021.
\newblock URL \url{https://www.R-project.org/}.

\bibitem[Reisach et~al.(2021)Reisach, Seiler, and Weichwald]{reisach2021beware}
Reisach, A., Seiler, C., and Weichwald, S.
\newblock Beware of the simulated {DAG}! {C}ausal discovery benchmarks may be
  easy to game.
\newblock In \emph{{A}dvances in {N}eural {I}nformation {P}rocessing {S}ystems
  ({NeurIPS})}, volume~34, 2021.

\bibitem[Rojas-Carulla et~al.(2018)Rojas-Carulla, Sch{\"o}lkopf, Turner, and
  Peters]{rojas2018invariant}
Rojas-Carulla, M., Sch{\"o}lkopf, B., Turner, R., and Peters, J.
\newblock Invariant models for causal transfer learning.
\newblock \emph{The Journal of Machine Learning Research}, 19\penalty0
  (1):\penalty0 1309--1342, 2018.

\bibitem[Schultheiss et~al.(2021)Schultheiss, B\"uhlmann, and
  Yuan]{schultheiss2021}
Schultheiss, C., B\"uhlmann, P., and Yuan, M.
\newblock Higher-order least squares: assessing partial goodness of fit of
  linear regression.
\newblock \emph{arXiv preprint arXiv:2109.14544}, 2021.

\bibitem[Shah \& Peters(2020)Shah and Peters]{shah2020hardness}
Shah, R.~D. and Peters, J.
\newblock The hardness of conditional independence testing and the generalised
  covariance measure.
\newblock \emph{Annals of Statistics}, 48\penalty0 (3):\penalty0 1514--1538,
  2020.

\bibitem[Shimizu et~al.(2006)Shimizu, Hoyer, Hyv{\"a}rinen, Kerminen, and
  Jordan]{shimizu2006linear}
Shimizu, S., Hoyer, P.~O., Hyv{\"a}rinen, A., Kerminen, A., and Jordan, M.
\newblock A linear non-{G}aussian acyclic model for causal discovery.
\newblock \emph{Journal of Machine Learning Research}, 7\penalty0 (10), 2006.

\bibitem[Spirtes et~al.(2000)Spirtes, Glymour, Scheines, and
  Heckerman]{spirtes2000causation}
Spirtes, P., Glymour, C.~N., Scheines, R., and Heckerman, D.
\newblock \emph{Causation, prediction, and search}.
\newblock MIT press, 2000.

\bibitem[Squires et~al.(2020)Squires, Wang, and Uhler]{squires2020permutation}
Squires, C., Wang, Y., and Uhler, C.
\newblock Permutation-based causal structure learning with unknown intervention
  targets.
\newblock In \emph{Conference on Uncertainty in Artificial Intelligence}, pp.\
  1039--1048. PMLR, 2020.

\bibitem[Takata(2010)]{Takata2010}
Takata, K.
\newblock Space-optimal, backtracking algorithms to list the minimal vertex
  separators of a graph.
\newblock \emph{Discrete Applied Mathematics}, 158:\penalty0 1660--1667, 2010.

\bibitem[Textor et~al.(2016)Textor, {van der Zander}, Gilthorpe, Liśkiewicz,
  and Ellison]{dagitty}
Textor, J., {van der Zander}, B., Gilthorpe, M.~S., Liśkiewicz, M., and
  Ellison, G.~T.
\newblock Robust causal inference using directed acyclic graphs: the {R}
  package `dagitty'.
\newblock \emph{International Journal of Epidemiology}, 45\penalty0
  (6):\penalty0 1887--1894, 2016.

\bibitem[Thams et~al.(2021)Thams, Saengkyongam, Pfister, and
  Peters]{thams2021statistical}
Thams, N., Saengkyongam, S., Pfister, N., and Peters, J.
\newblock Statistical testing under distributional shifts.
\newblock \emph{arXiv preprint arXiv:2105.10821}, 2021.

\bibitem[Tian et~al.(1998)Tian, Paz, and Pearl]{tian1998finding}
Tian, J., Paz, A., and Pearl, J.
\newblock Finding minimal d-separators.
\newblock Technical report, University of California, Los Angeles, 1998.

\bibitem[Tibshirani(1996)]{Tibshirani96}
Tibshirani, R.
\newblock Regression shrinkage and selection via the lasso.
\newblock \emph{Journal of the Royal Statistical Society, Series B},
  58:\penalty0 267--288, 1996.

\bibitem[van~der Zander et~al.(2019)van~der Zander, Li{\'s}kiewicz, and
  Textor]{van2019separators}
van~der Zander, B., Li{\'s}kiewicz, M., and Textor, J.
\newblock Separators and adjustment sets in causal graphs: Complete criteria
  and an algorithmic framework.
\newblock \emph{Artificial Intelligence}, 270:\penalty0 1--40, 2019.

\bibitem[Zhang et~al.(2011)Zhang, Peters, Janzing, and
  Sch{\"o}lkopf]{Zhang2011uai}
Zhang, K., Peters, J., Janzing, D., and Sch{\"o}lkopf, B.
\newblock Kernel-based conditional independence test and application in causal
  discovery.
\newblock In \emph{Proceedings of the 27th Annual Conference on {U}ncertainty
  in {A}rtificial {I}ntelligence ({UAI})}, pp.\  804--813, 2011.

\bibitem[Zheng et~al.(2018)Zheng, Aragam, Ravikumar, and Xing]{zheng2018dags}
Zheng, X., Aragam, B., Ravikumar, P.~K., and Xing, E.~P.
\newblock {DAG}s with {NO TEARS}: Continuous optimization for structure
  learning.
\newblock In \emph{{A}dvances in {N}eural {I}nformation {P}rocessing {S}ystems
  ({NeurIPS})}, volume~31, 2018.

\end{thebibliography}
\end{document}